\documentclass[aps,reprint,superscriptaddress, amsmath,amssymb,twocolumn, nobibnotes,nofootinbib,citeautoscript,floatfix]{revtex4-2}

\pdfoutput=1 
\setcitestyle{super}

\usepackage{float}
\usepackage{graphicx}
\usepackage{siunitx}
\usepackage{amsmath}
\usepackage{amssymb}
\usepackage{color}
\usepackage{soul}
\usepackage{braket}
\usepackage{commath,amsfonts, mathrsfs ,enumerate,booktabs}
\usepackage{color}
\usepackage{gensymb}

\usepackage{hyperref}

\newcommand{\icm}{$\mathrm{cm}^{-1}\,$}

\date{\today}
\begin{document}

\title{Retrieving genuine nonlinear Raman responses in ultrafast spectroscopy via deep learning}

\author{Giuseppe Fumero}
\thanks{Current affiliation: Associate, Physical Measurement Laboratory, National Institute of Standards and
Technology, Gaithersburg, MD, USA and Department of Physics and Astronomy, West Virginia University, Morgantown, WV, USA.}
\affiliation{Dipartimento di Fisica, Sapienza Università di Roma, Roma, Italy.}

\author{Giovanni Batignani}
\affiliation{Dipartimento di Fisica, Sapienza Università di Roma, Roma, Italy.}
\affiliation{Istituto Italiano di Tecnologia, Center for Life Nano Science @Sapienza, Roma, Italy.}

\author{Edoardo Cassetta}
\affiliation{Dipartimento di Fisica, Sapienza Università di Roma, Roma, Italy.}

\author{Carino Ferrante}
\affiliation{Dipartimento di Fisica, Sapienza Università di Roma, Roma,  Italy.}
\affiliation{CNR-SPIN, c/o Dipartimento di Scienze Fisiche e Chimiche, Università dell'Aquila, L’Aquila,  Italy.}

\author{Stefano Giagu}
\affiliation{Dipartimento di Fisica, Sapienza Università di Roma, Roma,  Italy.}

\author{Tullio Scopigno}
\affiliation{Dipartimento di Fisica, Sapienza Università di Roma, Roma,  Italy.}
\affiliation{Istituto Italiano di Tecnologia, Center for Life Nano Science @Sapienza, Roma, Italy.}
\affiliation{Istituto Italiano di Tecnologia, Graphene Labs, Genova, Italy.}

\begin{abstract}
\noindent Noise manifests ubiquitously in nonlinear spectroscopy, where multiple sources contribute to  experimental signals generating interrelated unwanted components, from random point-wise fluctuations to structured baseline signals. Mitigating strategies are usually heuristic, depending on subjective biases like the setting of parameters in data analysis algorithms and the removal order of the unwanted components. We propose a data-driven frequency-domain denoiser based on a convolutional neural network with kernels of different sizes acting in parallel to extract authentic vibrational features from nonlinear background in noisy spectroscopic raw data. We test our approach by retrieving asymmetric peaks in stimulated Raman spectroscopy (SRS), an ideal test-bed due to its intrinsic complex spectral features combined with a strong background signal. By using a theoretical perturbative toolbox, we efficiently train the network with simulated datasets resembling the statistical properties and lineshapes of the experimental spectra. The developed algorithm is successfully applied to experimental data to obtain noise- and background-free SRS spectra of organic molecules and prototypical heme proteins.
\end{abstract}

\maketitle
\renewcommand{\thefootnote}{\alph{footnote}}

Nonlinear optics has enabled and fostered the application of spectroscopy to the ultrashort time scales. Thanks to the developments of photonic techniques for femtosecond pulse generation and shaping, nonlinear spectroscopy addressed the interdisciplinary studies of ultrafast phenomena across a wide energy range \cite{Maiuri2019}, advancing, among others, the understanding of many-body interactions in structured and strongly correlated systems \cite{Han2022,Bloch2022,Mal2023}, carrier dynamics and e-ph couplings upon excitation \cite{Ulbricht2011,Zhu2018,Ferrante2022}, photochemical reactions, vibronic and non-adiabatic effects in molecules \cite{Musser2015,Neville2018,Zinchenko2021, Kuramochi2017, Fumero2020,Gaynor2019}. Similar advancements have also become possible in the closely related field of microscopy \cite{So2000,Min2011,Polli2018}. A common approach consists in using multiple ultrashort pulses, shaped in frequency and time, to resolve the induced modifications to a certain optical observable in a differential manner when one of the pulses is switched on and off. Dynamical insights can be obtained by adding a photochemical actinic pump and/or by tuning the pulses in resonance with the electronic absorption edges from which excited-state relaxation occurs. In particular, coherent resonant Raman spectroscopies exploit stimulated Raman scattering (SRS) to probe the structural response of the system undergoing ultrafast dynamics, thanks to their sensitivity to both the electronic and vibrational degrees of freedom \cite{Prince2016, Hall2017, Tahara2019, Ferrante2020,Fang2020,Malard2021}.
 
Even if nonlinearity is pivotal for accessing ultrafast dynamics over multiple time and energy scales, it is accompanied by two major complications in the data analysis and interpretation: 1) the signal generated by the nonlinear process is usually low compared to the residual, non-interacting light or to the fluctuations in the laser source, leading to point-wise noisy fluctuations in the measured spectra and 2) multiple nonlinear processes are usually generated by the same experimental layout, leading to the need of a-posteriori protocols to isolate the desired spectroscopic information from an unwanted baseline background and signal distortions due to overlaying competitive effects \cite{Hontani2018,Batignani2019,Ranjan2022,Genchi2023}. Post-processing routines are usually strongly dependent on the specific sample and on the experimental parameters. The sources of noise are not quantitatively known and often the exact lineshapes of both the baseline and signal cannot be predicted in advance \cite{Gautam2015,Shen2018}. Moreover, point-wise denoising and baseline subtraction are generally heavily correlated operations, which cannot be factorized, particularly when the target clean spectrum contains asymmetric lineshapes or components with largely different relative intensities \cite{Kloz2011}. The practical solution is often delegated to the experienced eye of the spectroscopist, a strategy which hampers automatization of the routine and may lead to sub-optimal resolutions, ambiguity and human biases. 

To overcome these critical issues and enhance the resolution of nonlinear Raman techniques beyond the limitations induced by the background and low signal-to-noise conditions, we have devised and trained a deep neural network (NN) based on multiple convolutional layers operating in parallel for denoising and baseline removal of raw SRS spectra. 
By designing the network architecture and choosing a suitable loss function and optimization strategy during training, we show how to perform the two tasks in parallel and avoid the difficulties which hinder the applications of standard data processing algorithms.

Applications of deep learning to nonlinear spectroscopy are still in their infancy but have already demonstrated great potential \cite{Thrift2020,Valensise2020,Hosseinizadeh2021, Valensise2021,Chen2021, Bresci2021, Stanfield2022, Vernuccio2022, Rankine2022}. NN have also been applied to preprocessing of spontaneous Raman data \cite{Wahl2020,Barton2021,Abdolghader2021,Gebrekidan2021,Shen2022,Luo2022}. The lineshapes in these techniques are always positive and the luminescence background is usually not structured. More importantly, the removal of the luminescent background and the denoising is tipically tackled separately. In the case of nonlinear Raman spectroscopy, due to asymmetric and complex lineshapes which are typical of SRS signals \cite{Batignani2016,Dobryakov2022,Batignani2022}, there are no optimal methods to disentangle the two tasks and perform them sequentially. Supervised training of the algorithms is further complicated by the absence of large enough labeled datasets ensuring a statistically relevant representation of the diverse SRS baseline and peak structures. To address these challenges, we combine here a multi-parallel convolutional NN architecture with a supervised training built on a theoretical toolbox based on the density matrix perturbative expansion for accurate modeling of the spectroscopic signals and their characteristic noise.
\section{Results}
\subsection{Signal and noise in Stimulated Raman Scattering}
\begin{figure*}[ht]
\centering
\fbox{\includegraphics[width=0.95\linewidth]{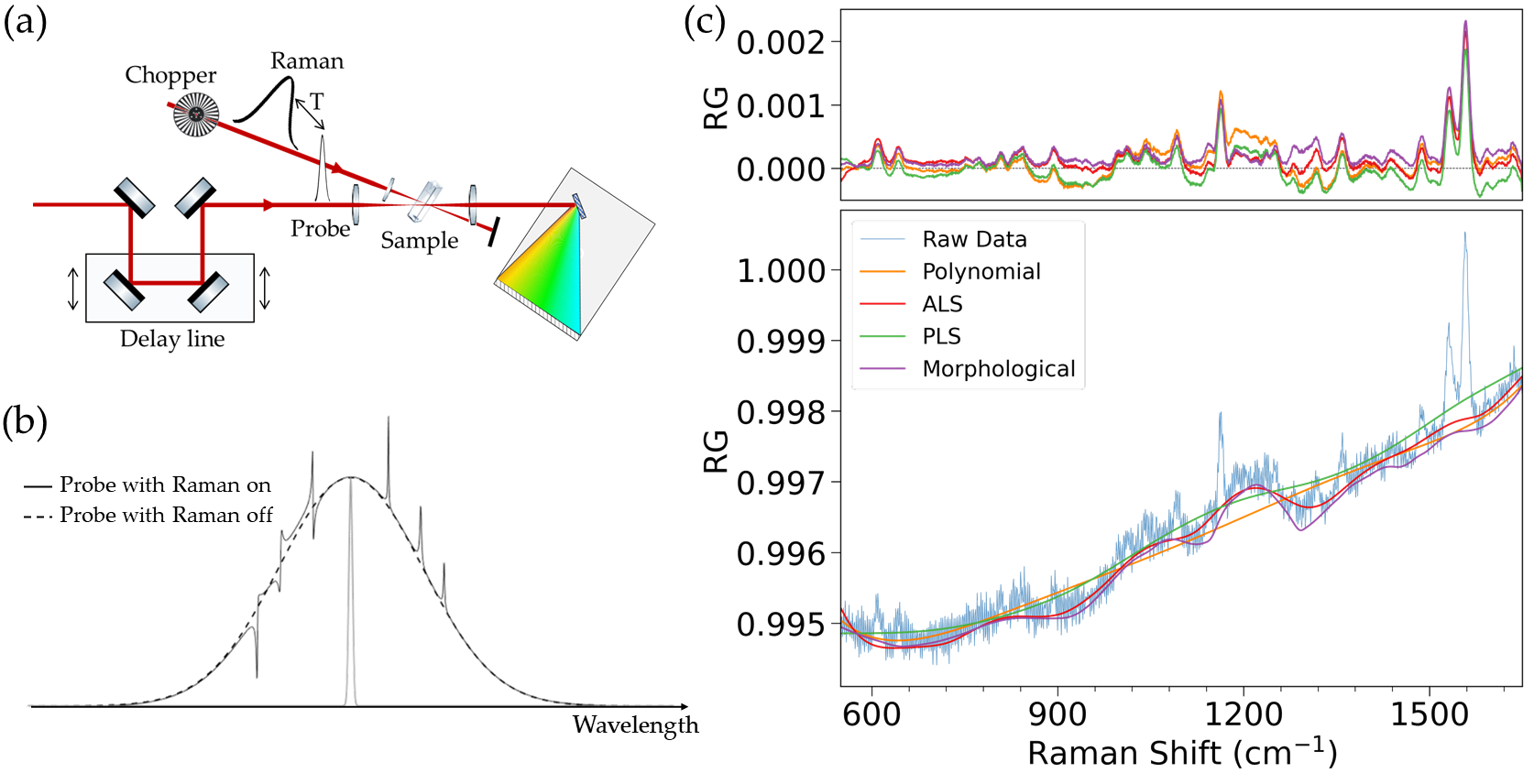}}
\caption{(a) Experimental scheme of SRS spectroscopy. A femtosecond broadband probe is focused on the sample together with a narrowband picosecond Raman pulse. The probe is then spectrally dispersed and detected. The relative arrival time of the pulses can be varied by a delay line. A mechanical chopper allows for the differential detection of the signal with and without the Raman pulse.  (b) Spectral envelope of the probe pulse detected after the sample in presence (black line) and in absence (black dashed line) of the Raman pulse (grey line). SRS features are obtained on the top of the probe spectrum when the two pulses interact with the sample.  (c) Example of the ambiguities that can arise from baseline removal in SRS red side data measured on a fluorescent protein (wt-GFP). The application of different algorithms - polynomial fitting, Asymmetric Least Square (ALS), Penalized Least Squares (PLS), iterative morphological fitting- determine the baseline differently (bottom panel). This results in differences in the retrieved SRS spectra (top panel). }
\label{fig:bsl}
\end{figure*}

SRS is a third order nonlinear optical effect that can be generated in the sample by the joint action of a broadband femtosecond Probe pulse (PP) and a narrowband picosecond Raman pulse (RP) overlapped in time \cite{Prince2016}, and then detected for spectroscopic applications to access the vibrational structure of the system under investigation (Fig. \ref{fig:bsl}a). The signal is coherently stimulated when the energy difference between the two laser pulses matches a Raman active molecular transition. Similarly to the spontaneous case, where the Raman peaks are spectrally located at the red and blue shifted sides of the excitation wavelength, the SRS signal is positively and negatively offset with respect to the RP. SRS features, however, arise as peaks, dips or even dispersive signals generated on top of the broadband spectral envelope of the probe, which is spectrally dispersed by means of a monochromator and detected (Fig. \ref{fig:bsl}b). SRS spectra usually display the normalized difference between the PP spectra recorded after the sample with the RP switched on and off (Raman Gain, RG), as a function of the detected frequency shift with respect to the RP (Raman shift). If the RP wavelength is tuned in resonance with the absorption of the sample, the transmission of the PP itself can be modulated even in the absence of stimulated Raman, due to the fast electronic response of the sample (Transient Absorption effect, TA) \cite{Berera2009}. This, together with additional nonlinear processes, including but not limited to solvent effects \cite{Roy2023} and nonlinear phase modulation \cite{Batignani2019}, causes the presence of an unwanted baseline, usually broader than the Raman features, which needs to be removed from the SRS raw data to extract the Raman spectra and correctly retrieve the vibrational information.

The treatment of raw SRS data are further complicated by phase and pulse-to-pulse instabilities of the laser sources and by the noise associated with the detection process \cite{Anderson2007,Audier2020,Robben2020}. This latter is typically due to the electronic fluctuations of the photogenerated carriers, the error related to the readout process, and the intrinsic shot noise limit caused by the quantum nature of light. Expected in any type of heterodyne-detected spectroscopy, these sources of noise combine with the inherent ambiguity of the baseline subtraction procedure, with the result that the overall signal-to-noise ratio achieved in real case scenarios is often considerably lower than the nominal sensitivity of the technique. This is exemplified in Fig. \ref{fig:bsl}c, in which we show the outcomes of different baseline estimation techniques applied to a typical SRS spectrum of the wild-type Green Fluorescent Protein. Without any assumption on the nonlinear dynamics and  on the Raman peaks lineshapes and spectral positions, the estimated baseline is highly dependent on the subtraction algorithm.  Consequently, it is often not possible to infer any conclusive statement on such ambiguously extracted features.

In order to simulate SRS spectra closely resembling the data obtained by the experiments, we used a perturbative framework based on the density matrix expansion \cite{Mukamel_book}. We simulated two datasets with different levels of noise, the \emph{high noise} (HN) and \emph{low noise} (LN) datasets. Each dataset consists of 5000 raw SRS spectra of 801  points each, with noise and baseline, associated to the corresponding clean spectra, serving as ground truth (GT). 80\% of the samples (4000 simulated spectra for each dataset) were used during the training phase. Additional details on the datasets are given in Methods and SI, section Ib.
\begin{figure*}[ht]
\centering
\fbox{\includegraphics[width=0.95\linewidth]{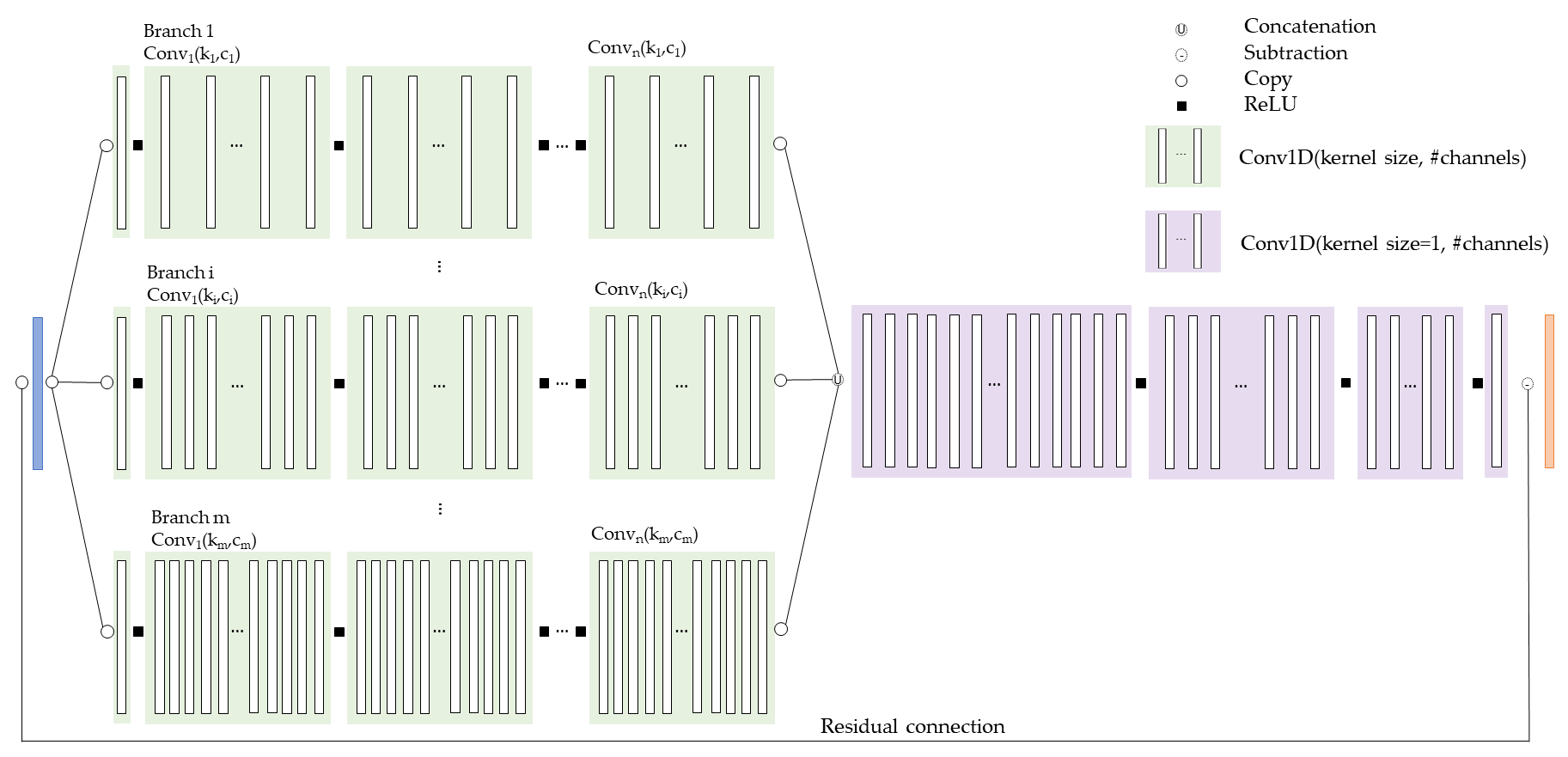}}
\caption{Architecture of the neural network. Input data (blue box) feed multiple convolutional branches which operates in parallel (green boxes). Within each branch, data are transformed by convolution layers ($\mathrm{Conv_1, \dots, Conv_n}$) with the same kernel size and number of channels, connected by nonlinear activation functions (ReLU, black squares). Concatenated output from the convolutional branches feed a series of 1D convolutional layers. The output (orange box) is obtained after a last residual layer. }
\label{fig:architecture}
\end{figure*}
\subsection{Supervised training of the neural network}
Considering the different spectral scales present in the problem, we adopted a convolutional neural network architecture \cite{LeCun1989,Zeiler2014,Zhang2017} combining different kernel sizes in parallel to reach different receptive fields at the same time. The general architecture of the model is shown in Fig. \ref{fig:architecture} and is based on parallel linear and nonlinear transformations consisting in zero padded convolutions with different kernel sizes and nonlinear activation functions. Each of the $N_{\mathrm{kernel}}$ parallel convolutional branches features $N_{\mathrm{layers}}$ layers of branch-depended kernels with $N_{param}$ trainable parameters. A detailed description is provided in Methods. The network's output is  $y=f_{\theta,w,b}	(x)$ where the parametric nonlinear map $f_{\theta,w,b}$ depends on the hyperparameters $\{\theta\}$, summarized in table \ref{tab:hyperparam}, and on the weights $w$ and biases $b$ of the convolutional layers. The task of the training step is to learn the set of $\{w,b\}$ which solve the minimization problem \cite{LeCun2015}
\begin{equation}
\underset{\{w,b\}}{\mathrm{argmin}} \sum_k L (f_{\theta,w,b}	(x_k),y^{GT}_k)
\label{eq: training}
\end{equation}
being the hyperparameters $\{\theta\}$ fixed before the training. In eq. \ref{eq: training},  $\{(x_1,y^{GT}_1),(x_2,y^{GT}_2),\dots (x_L,y^{GT}_L)\}$ is the training set of simulated noisy raw spectra ($x_k$) and corresponding clean ones ($y^{GT}_k$), i.e. the ground truth (GT). Both $x_k$ and $y^{GT}_k$ are 1D vectors of size $n_{input}$, functions of the sampled frequencies $\omega_1, \dots, \omega_{n_{input}}$. We performed stochastic gradient descent \cite{Ruder16} to minimize the custom loss function $L$:
\begin{equation}
\resizebox{.9\hsize}{!}{$L(y,y^{GT})=(1-W_{grad}) \Vert y-y^{GT} \Vert^2 +W_{grad} \mathcal{N} \Vert \nabla y - \nabla  y^{GT} \Vert^{\ell}$}
\label{eq: loss}
\end{equation}
where $\Vert \cdot \Vert^{\ell}$ indicates the $\ell$ norm and $y=f_{\theta,w,b}(x)$. $L$ contains a reconstruction term and a derivative term depending on the gradients of the reconstructed spectrum and GT. The two terms are activated at different epochs during training: initially only the reconstruction term is adopted for $N_{\mathrm{epoch}}^0$, allowing the algorithm to perform first estimations of possible baselines and noise without heavily impacting on the latter. Then the gradient term is included in the loss calculation and the model is trained for additional $N_{\mathrm{epoch}}^1$ epochs. The gradient term impacts heavily on the removal of the point-wise fluctuating noise, while the reconstruction term is sensitive to both the noise and corrections of the broadband baseline. The parallel optimization of these two terms is the key to obtain more accurate results with respect to performing these two tasks sequentially. The balance weight between the reconstruction and the gradient term is controlled by the hyperparameter $W_{grad} \in (0,1)$, which has been optimized by looking at the network performances and fixed to the value reported in table \ref{tab:hyperparam}. $\mathcal{N}$ is a normalization factor fixed once per training set only to regularize the loss decay during training, so that the value of the total loss does not experience an abrupt change during the first epoch after the gradient term is switched on.
\begin{table}[htbp]
\centering
\begin{tabular}{ccc}
\hline
     $\mathrm{\{\theta\}}$         & Network HN & Network LN \\ \hline
N\textsubscript{conv}   &     0       &    4         \\
N\textsubscript{param}  &     11k       &      10k       \\
N\textsubscript{kernel} &        63    &       21      \\
N\textsubscript{batch size} &        32    &       32      \\
$N_{\mathrm{epoch}}^0$ & 25   & 25 \\
$N_{\mathrm{epoch}}^1$        & 200   & 200 \\
$W_{grad}$                               & 0.6   & 0.6 \\
$\ell$        &     2       &       2      \\ \hline  
\end{tabular}
\caption{Hyperparameters for the HN and LN networks obtained by training with the HN and LN datasets and optimization with a grid search over the network architecture shown in Fig. \ref{fig:architecture}.}
\label{tab:hyperparam}
\end{table}
We found that the optimal architecture depends on the level of variability of the signal-to-noise in the training set. For a network trained on the dataset HN with high level of noise, the best architecture (network HN) was obtained for the value of the hyperparameters reported in the second column of table \ref{tab:hyperparam}. For the LN training set with a lower maximum value of noise, we obtained the best performances with the network specified by the hyperparameters in the third column of table \ref{tab:hyperparam} (network LN). The choice of the best hyperparameters to be used on experimental data can be performed using a discriminator on the signal-to-noise level of the data, as detailed in section II.d of the SI. 

\subsection{Validation on simulated data}
\begin{figure*}[ht]
\centering
\fbox{\includegraphics[width=0.95\linewidth]{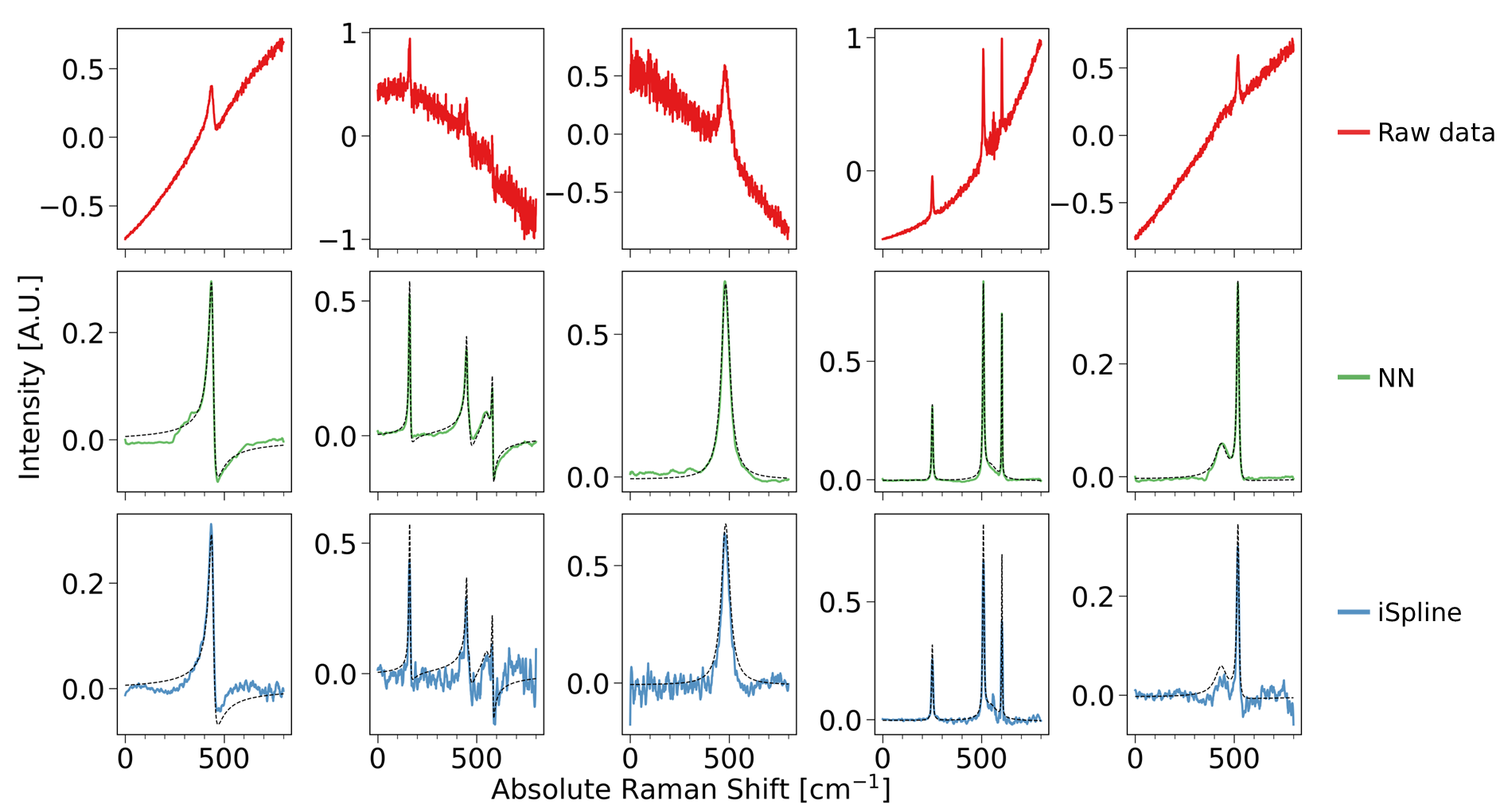}}
\caption{Evaluation of representative simulated test samples from the HN dataset. For each sample, noisy raw data are shown in the top panel (red lines). Central and bottom panels show the SRS spectra obtained by using the NN (green) or the iterative spline procedure for baseline subtraction followed by a Savitzky-Golay filtering (blue). GT are also reported for comparison (black dotted lines).}
\label{fig:simulated_spectra}
\end{figure*}

We tested the networks trained on the HN and LN datasets on test sets containing the 20\% of the samples of the corresponding datasets, previously unseen by the networks during training. In Fig. \ref{fig:simulated_spectra}, we present the result for five typical spectra extracted from the HN test set. The corresponding results and the analysis for the LN test dataset are reported in section II.c of the SI. The spectra reconstructed by the NN (green lines) show an excellent agreement with the clean GT spectra (black dotted lines), for very different baseline shapes, which are shown by the raw data in red in the top panel. These examples are compared to the results obtained by an iterative spline (iSpline) procedure for baseline removal \cite{Ferrante2018} followed by a Savitzky–Golay (SG) filtering. We note that the parameters of the SG filter have been chosen a posteriori by using the GT, in order to obtain a trade-off between an optimal smoothing of the point-wise noise and the preservation of the linewidth of the peaks. Consequently, this is an optimal procedure which cannot be meet in a real case scenario, i.e. when the GT is not available. Notwithstanding, the reconstruction of the lineshape, linewidth and relative intensity of the peaks and the overall removal of the baseline and the noisy point-wise fluctuations are qualitatively better when using the NN.	 

For a quantitative comparison, it is necessary to define indicators that measure the capability of a given algorithm to perform a specific task. We combined selected metrics from the literature with custom-defined ones to evaluate the performance of the NN for the tasks more closely related to spectroscopic purposes, namely the identification of peak position and lineshapes, evaluation of relative intensities, separation of overlapped components and minimization of false-peak predictions. Specifically, we computed the Structural Similarity Metric (SSIM), Normalized Mean Absolute and Mean Squared Errors (NMAE and NMSE), as defined in Methods. In addition to these standard metrics, we developed a custom edge finder, which determines the spectral positions of the positive and negative peaks from an input spectrum resulting from the processing of the noisy raw data and compares them to the associated GT spectrum. We used the edge finder to count the correctly assigned edges (true positive, TP) and the errors induced by the data processing (false positive and false negative, FP and FN). The edges are considered TP if their spectral position deviates from the corresponding GT more than a tunable value of tolerance. In the following, we set the tolerance to one pixel. This allows us to redefine this problem to a classification task and calculate standard metrics as the \emph{F1 score} and \emph{precision} on retrieving the spectral positions of the features accurately. Finally, we defined a custom Signal-to-Noise Ratio (SNR) metrics as
\begin{align}
\begin{split}
\mathrm{SNR} &= \frac{Area(I_y)-\langle N_y \rangle }{\Vert N_y-N_{y^{GT}} \Vert^2} \\
I &= y(\omega_{\nu}) \\
N_f &= f(\omega\neq \omega_{\nu}) \text{ with } f=y,\,y^{GT}
\end{split}
\end{align}
where $\omega_{\nu}=\{\omega_i, \quad \forall i: \lvert \omega_i-\omega_{peak}^{GT}\rvert<\epsilon \}$ and $\epsilon=80$ pixels, which was set taking into consideration the typical width of the Raman features.
This metric measures the capability of an algorithm to obtain a clean and smooth baseline with respect to the area of the main peaks in the spectral position defined by the GT. 
\begin{figure*}[ht]
\centering
\fbox{\includegraphics[width=0.95\linewidth]{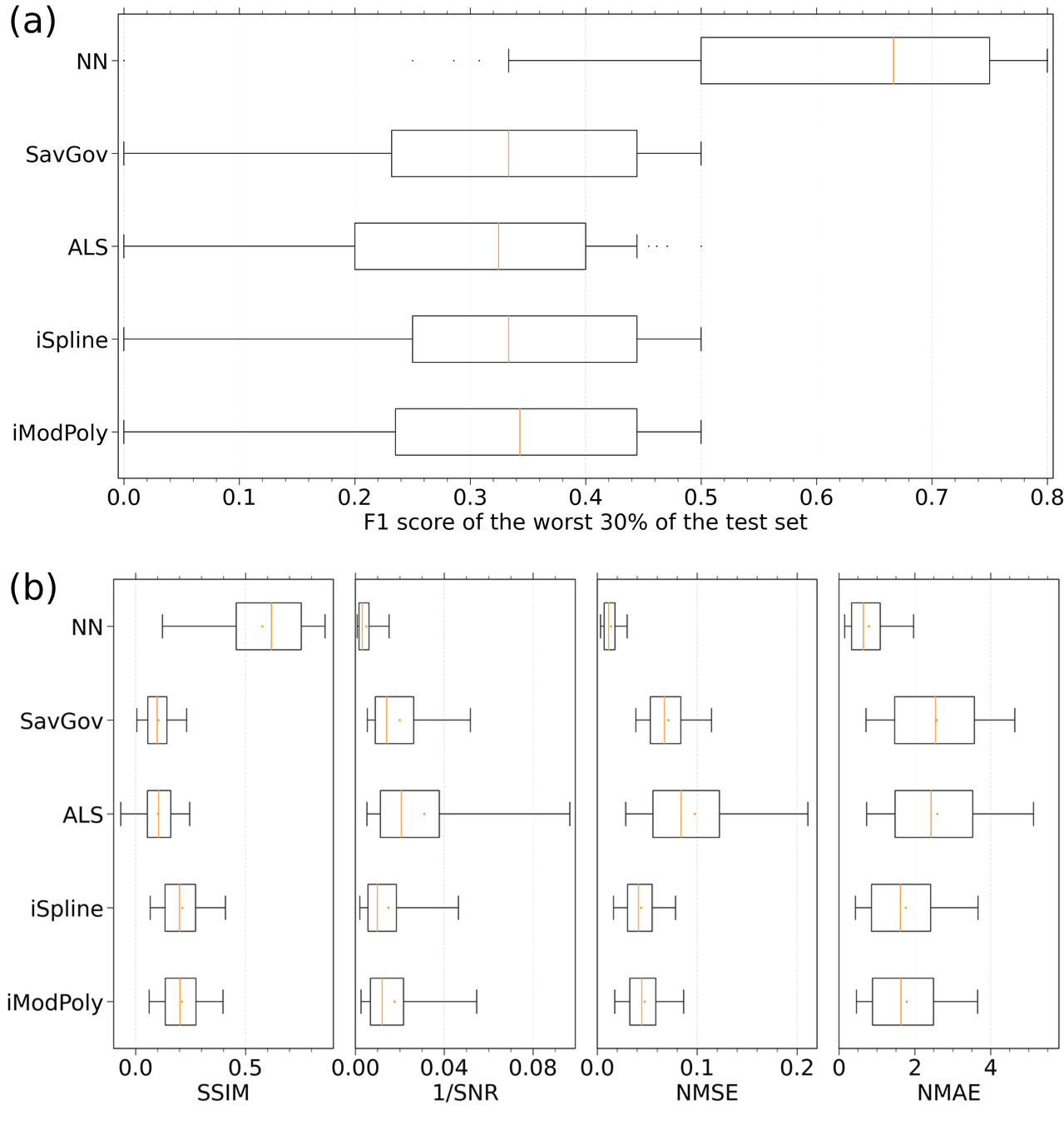}}
\caption{Comparison of the statistical analysis over selected metrics between the NN method and multiple non-data driven routines applied to the HN test set. Panel a shows the results of the classification metric obtained by means of the edge finder algorithm. The whisker plots report on the F1 score relative to the portion of 30\% of the test set for which each method has the worst score. Panel b shows the whisker plots relative to the results obtained by the different methods measure by the SSIM, SNR, NMSE and NMAE metrics. For all the whisker plots, the box covers from first to the third quartile, while the whiskers extend from the box to the 5th and to the 95th percentile. The orange line and dot indicate the median and the mean, respectively. Black dots indicate values that are past the end of the whiskers.}
\label{fig:metrics}
\end{figure*}

We have compared the results obtained by the NN with those obtained by four different traditional algorithms for baseline removal: third-order modified polynomial (iModPoly) \cite{Zhao2007}, iterative spline, SG filter and peak-screened Asymmetric Least Squares (ALS) \cite{Korepanov2020}. For each algorithm,  baseline removal was followed by an additional SG filter for point-wise denoising and smoothing (see Methods for additional details). The results obtained for the HN dataset are shown in Fig. \ref{fig:metrics}, while the corresponding results for the LN set are reported in section II.c of the SI. The NN outperforms the standard algorithms in identifying all the edges in the SRS spectrum, with a full precision score achieved in the $\mathrm{63\%}$ of the test samples, compared to the $\mathrm{42\%}$ of the best standard algorithm (iSpline+SG filter). More importantly, the distribution of the F1 score achieved by the NN is narrower and shifted to the higher values, with a mean of $0.86$ and standard deviation of $0.18$ to be compared with the mean of  $0.70$ and standard deviation of $0.30$ of the standard data processing routine. This is shown in Fig. \ref{fig:metrics}a, where the distribution of the results on the F1 score for the worst quartile of the test set is presented by whisker box plots, while the full histogram is reported in Fig. S2 of the SI. We note that the values of precision depend on the amount of tolerance on the peak position discrepancy between the GT and reconstructed data, but this dependence does not alter the results presented here (details are reported in SI, section IIb). Also for the other tasks that we have considered, the NN shows superior performances, as illustrated by the analysis of the associated metrics in Fig. \ref{fig:metrics}b. In particular, for the metrics NMAE, NMSE and for 1/SNR, which project optimal results towards zero, the medians  relative to the NN are lower than the first quartile of all the distributions of the standard methods. There is also a large improvement in  the SSIM metric - optimal at $\mathrm{SSIM=+1}$ - for which the median of the NN is larger than the 95th percentile of the best standard method. This is correlated to the ability of the NN to also recover the overall lineshape of the Raman features, in addition to the signal-to-noise, intensity and peak positions, which are best measured by the other metrics. The performances of the standard methods are comparable among each other, with the polynomial and iterative spline algorithms giving the best results. 

\subsection{Validation on experimental data}

\begin{figure*}[ht]
\centering
\fbox{\includegraphics[width=0.95\linewidth]{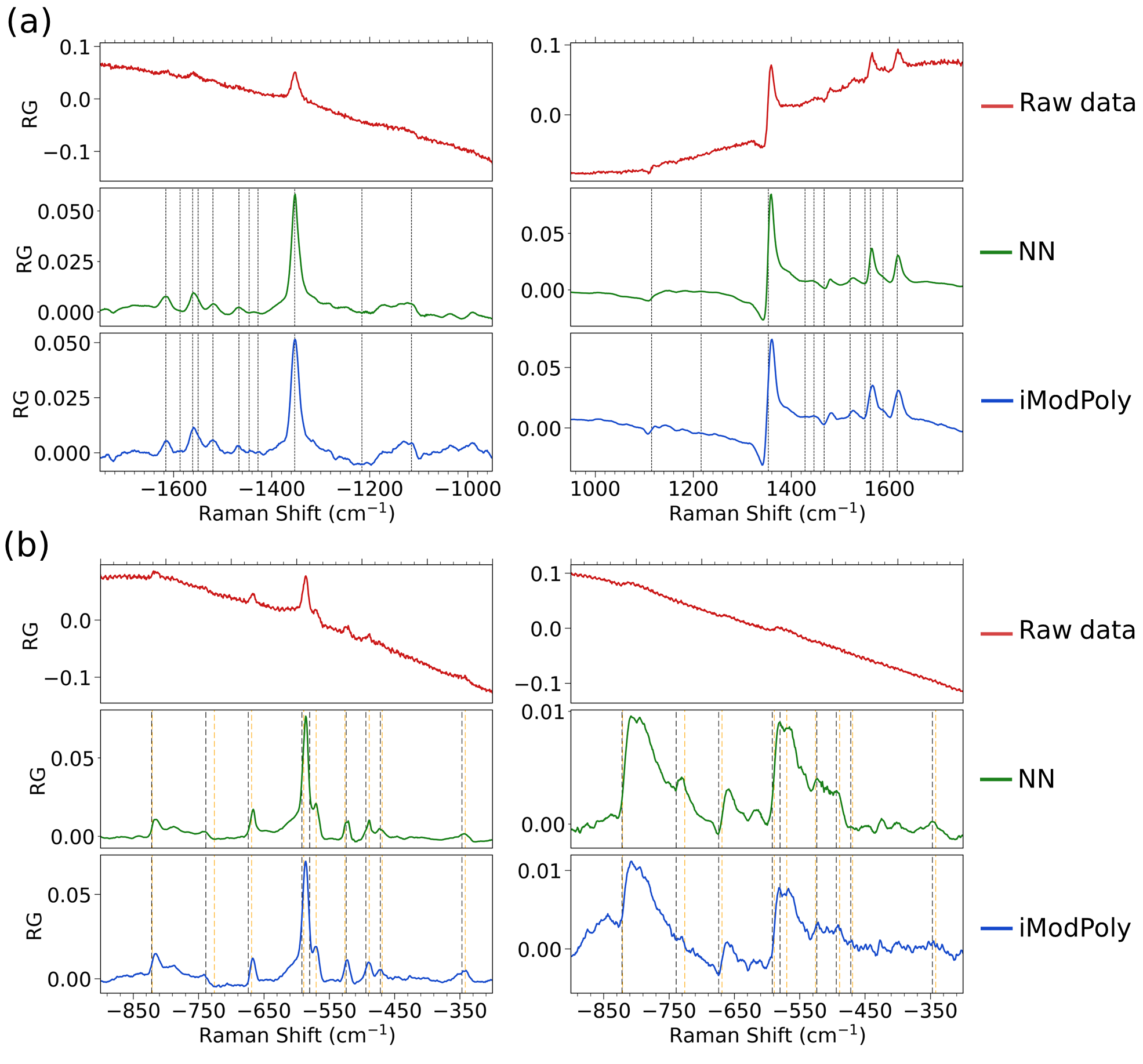}}
\caption{Application to SRS experimental data. (a) SRS spectra of deoxy Mb. Left panels: red side Raman spectra obtained with RP tuned at 447 nm. Right Panels: blue side Raman spectra obtained with the RP tuned at 472 nm. Raw data are indicated by red lines, while green and blue lines represent processed spectra predicted by the LN network and by the iModPoly algorithm for baseline removal followed by a SG noise filter. Vertical dashed lines indicate the spectral position of the Raman modes of deoxy Mb for resonant excitation in the Soret absorption band \cite{Ferrante2018} (reported in SI, table S2). (b) SRS spectra of Cresyl Violet obtained upon resonant excitation with a RP tuned at 580 nm at two different delays between the RP and PP: $\mathrm{-0.8}$ ps (PP preceding RP, left panels) and $\mathrm{+1.2}$ ps (PP following RP, right panels). Raw data are indicated by red lines, while green and blue lines represent processed spectra predicted by the LN network and by the iModPoly algorithm for baseline removal followed by a SG noise filter. Vertical dashed lines indicate the spectral position of the Raman modes of the ground (in black) \cite{Fitzpatrick2020,Batignani2020,Lu2020, Batignani2021} and the excited states (in orange) \cite{Fitzpatrick2020,Batignani2020,Lu2020}, which are reported in SI, table S3.}
\label{fig:exp_spectra}
\end{figure*}

To validate the NN denoiser in a real case scenario, we have applied it to the resonant SRS spectra measured on equine heart deoxy Myoglobin (Mb) dissolved in pH 7.4 buffer and Cresyl Violet (CV) dissolved in methanol, using the hyperparameters found by training with the LN dataset, whose signal-to-noise ratios are comparable to the ones obtained for these particular samples with our SRS setup. In Fig. \ref{fig:exp_spectra}a, we report the raw experimental SRS spectra of Mb pumped across the Soret absorption band, with the RP wavelength tuned at $\mathrm{447}$ nm and $\mathrm{472}$ nm for the red and blue sides of the spectrum, respectively. The resonant condition impacts differently on the two sides of the spectrum and makes the Raman lineshapes highly dependent on the RP wavelength and, for the blue side, on the frequency of the normal mode \cite{Batignani2016}. In the central and lower panels, we show the corresponding spectra processed by the NN and the polynomial algorithm. The results are in good agreement with the literature \cite{Spiro11988,Ferrante2018} and the best results obtained by traditional processing routines, for both the gain and dispersive lineshapes.

We then tested the algorithm in the more demanding case of CV undergoing photoexcitation due to a resonant, high-fluence RP tuned at 580 nm. In these conditions, the RP also acts as an actinic pump, inducing electronically excited-state populations in the sample by means of two additional light-matter interactions preceding the SRS process \cite{Batignani2020}. The efficiency of this higher-order nonlinear process depends on the time delay between the Raman and the probe. Tuning the delay, it is possible to control the amount of excited-state population induced by the Raman pulse. Here a positive (negative) sign indicates that the PP follows (precedes) the RP. At large negative delays, the Raman scattering mainly involves molecules initially in the electronically ground state. At positive delays, SRS can probe vibrational transitions originating from a mixture of excited and ground-state populations, with the cost of a lower signal-to-noise ratio, spectral distortions and an overall increased complexity of the baseline and lineshapes. We considered the SRS spectra for two different delays between the Raman and probe pulses: $\mathrm{-0.8}$ ps (PP preceding RP) and $\mathrm{+1.2}$ ps (PP following RP). In Fig. \ref{fig:exp_spectra}b we report the raw SRS spectra at the two delays and the corresponding ones processed by the NN and by the iModPoly algorithm followed by SG filtering. For the negative delay (left panel), the processed spectra present sharp and intense features, with a high signal-to-noise ratio and spectral positions matching the frequencies of both the electronically excited and ground-state Raman modes of CV, due to the resonant RP wavelength. For the positive delay (right panel), as a consequence of the interference with the additional Raman processes originating from the excited-state populations, spectral features appear broadened and red-shifted, their lineshapes become dispersive and asymmetric and the signal-to-noise ratio decreases drastically, with different impact on the two methods used to process the raw spectra. Notably, the combined iModPoly and SG algorithm are not able to resolve the excited-state peak at 733 \icm, which is conversely retrieved by the NN. The polynomial processing also shows parasitic broad peaks between 850 and 900 \icm, and poorly captures the dispersive lineshape at 680 \icm. Moreover, the two approaches give different relative intensities between the peaks, as also observed for the simulated datasets. These results confirm that the NN algorithm achieves higher performances also in the case of complex experimental spectra. 
\section{Discussion}
Low signal-to-noise ratios and structured spectral lineshapes in nonlinear Raman spectroscopy cause loss of information and increase the complexity in performing the measurements, due to the need of longer exposure times for averaging, and data analysis. We have demonstrated that wisely-designed deep neural networks can overcome these limitations and achieve background removal and denoising of stimulated Raman spectra. By means of multiple kernel sizes operating in parallel within a convolutional residual neural architecture, it is possible to adapt the receptive field of the network to the informative features in the spectra and treat the multiple spectral scales present in the SRS data, related to the complex Raman lineshapes, to the background and to point-wise noisy fluctuations. We have shown that such architecture is able to adapt to different levels of noise and prominence of the Raman bands from the baseline, by training on datasets simulated through the diagrammatic theory and characterized by noise and material parameters resembling those present in experimental conditions. Depending on the level of noise, the NN demonstrated itself comparable or superior to the standard algorithms commonly used for SRS data postprocessing. The advantages are particularly evident in presence of multiple features with asymmetric lineshapes and intensities that are weak compared to the noise and to the other bands in the same spectral region. The NN algorithm was able to identify the Raman bands, reconstruct the correct lineshapes and relative intensities and enhance the spectral resolution by resolving the vibrational frequencies and bandwidths for close or overlapping Raman features. These abilities are pivotal for the interpretation of experiments leveraging on resonant optical excitations to detect cooperative mechanisms between coherent vibrations and electronic excitations. Once trained, the network generalizes to experimental spectra obtained on different samples, preserving its high performances. We note that SRS is a optimal test-bed for investigating deep learning applications to nonlinear spectroscopy, given its spectral complexity and variety of lineshapes for both the Raman features and background contributions. For these reasons, we anticipate the possibility to extend the use of the proposed NN architecture to different linear and nonlinear spectroscopic techniques which are affected by similar noise effects by means of fine tuning of the final NC layers and transfer learning techniques.
\section{Methods}
\subsection{Simulated training datasets}
The HN and LN datasets have been simulated using the nonlinear response perturbation theory. In this theoretical framework, the nonlinear signal is obtained from the $n$-th order nonlinear optical polarization $P^{(n)}$, which consists of the convolution between matter correlation functions and the electromagnetic fields. The radiation-matter interaction is treated perturbatively and the density matrix is expanded in power of the fields, applying many-body Green function techniques in the Liouville space. Diagrammatic representations are exploited to isolate all the relevant terms in the expansion and calculate non-equilibrium expectation values of the correlation functions \cite{Dorfman2013,Fumero2015}. Different sets of diagrams are associated to the SRS signal and to the TA baseline \cite{Kowalewski_Mukamel2017}, which can be simulated separately (see SI, section Ia). Shot noise is also included by means of fluctuations scaling as the square root of the number of detected photons and the associated uncertainty is propagated through the spectroscopic signal\footnote{The electronic noise due to the detection process is normally negligible with respect to the shot noise when using single shot detection with a large dynamic range CCD camera, which is the common acquisition scheme for SRS. In different contexts, it can be included by means of a random additive noise.}. Consequently, all the three different spectral scales which are peculiar of the SRS data are present in the simulations: long-scale baseline variations, point-wise noise fluctuations and, in between of these two extrema, the Raman features. Within each dataset, half of the samples has been generated by simulated signals in the red side of the SRS spectrum with respect of the narrowband RP, half in the blue side. All the molecular and experimental parameters have been varied randomly through the datasets and sampled from a uniform distribution within the boundaries summarized in the SI, table S1. The main difference between the two datasets consists in the different number of averaged acquisitions per sample, $\mathrm{N_{acquisition}}$. Averaging impacts on the overall noise of the sample as a multiplicative factor inversely proportional to the squared root of the number of acquisitions. Each sample in a dataset resulted from the average of $\mathrm{N_{acquisition}}$ noisy replica of the experiment. $\mathrm{N_{acquisition}}$ was set to 1 and to 100 for the HN and LN datasets, respectively. To avoid biases due to the selection of a particular dynamical model or pulses synthesis methods, no explicit dynamics is considered and the pulses are modeled as Gaussian temporal profiles with random duration. A variety of possible lineshapes and spectral properties is sampled by the randomly selecting the experimental and molecular parameters.
\subsection{Neural network architecture}
The network takes as an input the SRS spectra sampled at $n_{input}$ points in frequency. The input is replicated to feed $N_{\mathrm{kernel}}$ parallel branches, which are composed by $N_{\mathrm{layers}}$ convolutional blocks each and return feature maps of different dimensions due to the different number of output channels. In particular, each block performs $c_i$ convolutions with a different, randomly initialized filter of size $k_i$, followed by a Rectified Linear Unit (ReLU) activation function \cite{Goodfellow2016}. Both the number of filters $c_i$ and the kernel size $k_i$ are branch-dependent and fixed along the branch. After the last convolution block, each branch returns a feature map of dimension $b+\{c_i,(1,n_{input})\}$ where $b$ is the size of the training batches. The features from different branches are concatenated along the channel dimension after the last convolutional block of each branch and feed a single final branch of $N_{\mathrm{conv}}$ convolutional blocks with ReLU activation. In the final branch the kernel size is fixed to one and the number of filters downscales by half each block. In such a way, we obtain a parametric linear combination of the features obtained from the different branches to be optimized by training. The last convolution has $n_{input}$ filters so that the output channel dimension matches the input to recover feature interpretability and precedes a final residual step in which the input is subtracted from the feature map. The residual layer allows the neural model to learn only the structure of the noise, reducing the number of required labeled samples and the overall complexity of the training phase. The final model architecture was selected after cross-validating over all the  model hyperparameters by means of grid-search, including $N_{\mathrm{layers}}$, $N_{\mathrm{conv}}$, the  kernel sizes, learning rate, batch sizes, number of training epochs and the relative weight between the gradient and reconstruction loss terms. The learning rate was also dynamically decreased during training accordingly to a scheduler. This also allowed for balancing the effects of the training with the full loss function in eq. \ref{eq: loss}, which occurs with a smaller learning rate. Batch normalization and regularization techniques were not adopted since overfitting was not observed and they decreased the performance of the neural net. The kernel sizes have been varied linearly across the branches, from a minimal size of 5 to a maximal size $M_{k}$ which was chosen by an hyperparameter sweep and then fixed to $M_{k}=88$. The number of filters and hence the output channels of each convolutional block was chosen in a kernel-size dependent way to obtain the same number of parameters $N_{param}$ in each convolutional branch. Also $N_{param}$ has been chosen by an hyperparameter sweep, in order to find a trade-off between the needs of computational feasibility and performance. Backpropagation with the Adam optimizer was used to train the model \cite{Goodfellow2016}.
\subsection{Algorithm implementation}
The algorithms described in this work have been developed in Python (3.7) based on the open source libraries Keras and TensorFlow (2.10) for deep learning and the NVIDIA driver CUDA (10.1) for GPU acceleration. Training was performed on a Tesla V100-SXM2-32GB provided by the dual NVIDIA DGX-1 at SapienzaAI \cite{SapienzaAI}. Training and optimization have been tracked by using the Weights\&Biases platform \cite{wandb}. All the samples in each dataset have been preprocessed by subtracting the mean and normalizing by the standard deviation calculated across the dataset and further rescaled by the maximum of the training set. The experimental data have been preprocessed in the same way and then rescaled by the same factor used for the training data. The optimal learning rates to be used during the training have been identified during the hyperparameter sweep and then fixed. The initial learning rate was set to $10^{-3}$ and then  to $5\cdot 10^{-4}$ at the beginning of the second phase of the training, after $N_{\mathrm{epoch}}^0$ epochs, when the gradient term is included in the loss function. Additionally, the value of the learning rate has been slightly decreased every epoch by means of an exponential scheduler. The final model required 1h46m to be trained for the HN network and 35 minutes for LN network. Code with minimal example implementation, datasets and pretrained weights discussed here are available online \cite{SRSdenoiser_rep}.

\subsection{Standard denoising and background subtraction methods}
In this work, we compared four standard methods for baseline removal:
\begin{itemize}
\item Savitzky-Golay (SavGov)
\item Asymmetric Least Square (ALS)
\item iterative Spline (iSpline)
\item improved Modified Multi-Polynomial (iModPoly)
\end{itemize}
We combined each method with a sequential application of a Savitzky-Golay filter \cite{Savitzky1964} for point-wise noise subtraction. For each spectral bin of a sample in the dataset, the SavGov filter uses a least-squares procedure to fit the data in a window of length $l$ centered at the input bin to a polynomial of order $p$ and  replaces the input value with the central point of the fitted polynomial function. We used the SciPy implementation \cite{2020SciPy-NMeth} and fixed the order of the polynomial to $p=3$ and the window length to $l=551$  for the baseline removal routine and to $l=17$ for the denoising filter. 
In the iSpline method, we used iterative spline interpolation  to distinguish the baseline and the Raman features, thanks to their different spectral complexity. Specifically, initially the spline is performed on the full spectral range, and the regions that deviate from it up to a positive fixed threshold are defined as Raman-like. Then, the interpolation is repeated for the spectral regions marked as baseline in the previous step with a lower threshold, until the number of iterations provided by the user is reached. We chose the thresholds to be proportional to the standard deviation of the retrieved baseline.
The ALS algorithm attempts to find the baseline by a weighted least square fit in which  values of the candidate baseline lower than the data at a given point are favored and in which the second derivative is penalized at the same time. We used a modified version implemented in \cite{Korepanov2020}, which includes peak screening.
Finally, the iModPoly baseline subtraction routine \cite{Zhao2007} consists in an iterative polynomial fitting with a peak-removal procedure and a correction to account for noisy fluctuations by means of the standard deviation of the residual. For each iteration, to classify a specific region of the data as either Raman feature or baseline, the input is compared to the sum of the residuals of the fit and their standard deviation. We used the implementation in \cite{PyPI} with a third order polynomial. 
\subsection{Evaluation metrics}
The SSIM reported in the Results section measures the structural resemblance of two dataset within a given convolution window by means of the following quantity:
\begin{equation}
    \mathrm{SSIM}(x,y)= l(x,y)c(x,y)s(x,y)
\end{equation}
with
\begin{align}
\begin{split}
    &l(x,y)= {\frac{2\mu_x\mu_y+C_1}{\mu_x^2+\mu_y^2+C_1^2}}\\
    &c(x,y)= {\frac{2\sigma_x\sigma_y+C_2}{\sigma_x^2+\sigma_y^2+C_2^2}}\\
    &s(x,y)= {\frac{\sigma_{xy}+C_3}{\sigma_x\sigma_y+C_3^2}}
\end{split}
\end{align}
and $\mu_i$,$\sigma_i$ and $\sigma_{ij}$ for $i,j=x,y$ are the local mean, standard deviation and covariance of the two datasets within the convolutional window, $C_n=K_n L$ for $n=1,\dots,3$, being $L$ the dynamic range of the data, are regularizers to avoid singularities. Following \cite{Wang2004}, we set $C_1=0.01$, $C_2=0.03$ and $C_3=\frac{C_2}{\sqrt{2}}$.
Mean SSIM is then calculated as the mean over all the scanned windows. In addiction to the signal-to-noise-ratio (SNR) defined in the main text, we have also analyzed Normalized Mean Absolute Error (NMAE)
\begin{equation}
\mathrm{NMAE} = \frac{\Vert y-y_{GT} \Vert}{\Vert y \Vert} 
\end{equation}
and Mean Squared Error (NMSE)
\begin{equation}
\mathrm{NMSE} = \frac{\Vert y-y_{GT} \Vert^2}{ \Vert y \Vert^2}  
\end{equation}
which are less and more sensitive measures to outliers, respectively. Finally, using the custom edge finder \cite{SRSdenoiser_rep}, we redefined the evaluation problem as a classification task and calculated the values of true positives (TP), false positives (FP) and false negatives (FN) predicted by the NN or the reference standard algorithm with respect to the GT. Here with edge we refer to both gains and losses in the Raman spectra. From these values, standard metrics as the \emph{precision} and \emph{recall} have been calculated:
\begin{equation}
\mathrm{precision} = \frac{TP}{TP+FP}
\end{equation}
\begin{equation}
\mathrm{recall} = \frac{TP}{TP+FN}
\end{equation}
Precision measures the number of correctly retrieved edges scaled by the total number of edges found, while recall scales the retrieved true positives to the number of all the actual positive values (TP+FN). The F1-score is calculated as the harmonic mean of precision and recall and is a trade-off between possible predicting biases of a model towards high false positives and false negatives:
\begin{equation}
\mathrm{F1} = 2 \,\frac{\mathrm{Precision}\cdot\mathrm{Recall}}{\mathrm{Precision}+\mathrm{Recall}}= \frac{2TP}{2TP+FP+FN}
\end{equation}
\subsection{Spectroscopic setup and measurements}
The SRS setup have been described in details in previous works \cite{Ferrante2018}. The two beams needed for SRS are synthesized from a common regenerative amplified Ti:sapphire mode-locked laser, which generates 3 mJ, 40 fs pulses at 800 nm with 1 kHz of repetition rate. The femtosecond probe pulse is obtained by white light continuum generation  in 2-mm-thick Sapphire crystal.  A narrowband Raman pulse is synthesized and tuned by means of a two-stage optical parametric amplifier (OPA) followed by a spectral compression stage based on frequency doubling in a 25 mm beta barium borate crystal. An additional cleaning of the spectral frequency is performed by a double-pass (2f) spectral filter. The temporal ordering of the pulses is controlled by mechanical delay lines. The Raman and probe pulses have linear and parallel polarization. After interaction with the sample the probe pulse is frequency dispersed by a spectrometer onto a CCD, able to perform single-shot acquisitions. A synchronized chopper blocks alternating Raman pulses in order to obtain the Raman gain using successive probe pulses. Cresyl Violet (CV)  was dissolved in methanol and measured at ambient temperature. Details on the preparation of the Mb can be found in \cite{Ferrante2018}.

\section*{Data and code availability}
The codes, datasets and minimal implementation examples relevant to this work are available at \url{https://github.com/Gifum/SRSdenoiser}. \\

\section*{Supporting information}
Supporting information are available for i) Generation of the simulated datasets, ii) Analysis of the performance of HN and LN networks on the simulated datasets and iii) Frequencies of the Raman modes in the experimental spectra.

\begin{acknowledgments}
\noindent G.F. would like to thank Marco Fumero for many fruitful discussions. This project has received funding from the PRIN 2020 Project, Grant No. 2020HTSXMA-PSIMOVIE  (G.B.), from the European Union’s Horizon 2020 research and innovation program Graphene Flagship under Grant Agreement No. 881603  (T.S.) and has been partially supported by PNRR MUR project PE0000013-FAIR (S.G.). G.B. and T.S. acknowledge the `Progetti di Ricerca Medi 2020', the `Progetti di Ricerca Medi 2021', the `Progetti di Ricerca Medi 2022' grants by Sapienza~Università~di~Roma. G.F. acknowledges support from the grant `Avvio alla Ricerca 2022' by Sapienza Università di Roma.
\end{acknowledgments}

%
\end{document}


\title{Supplementary Information: Retrieving genuine nonlinear Raman responses in ultrafast spectroscopy via deep learning}

\author{Giuseppe Fumero}
\thanks{Current affiliation: Associate, Physical Measurement Laboratory, National Institute of Standards and
Technology, Gaithersburg, MD, USA and Department of Physics and Astronomy, West Virginia University, Morgantown, WV, USA.}
\affiliation{Dipartimento di Fisica, Sapienza Università di Roma, Roma, Italy.}

\author{Giovanni Batignani}
\affiliation{Dipartimento di Fisica, Sapienza Università di Roma, Roma, Italy.}
\affiliation{Istituto Italiano di Tecnologia, Center for Life Nano Science @Sapienza, Roma, Italy.}

\author{Edoardo Cassetta}
\affiliation{Dipartimento di Fisica, Sapienza Università di Roma, Roma, Italy.}

\author{Carino Ferrante}
\affiliation{Dipartimento di Fisica, Sapienza Università di Roma, Roma,  Italy.}
\affiliation{CNR-SPIN, c/o Dipartimento di Scienze Fisiche e Chimiche, Università dell'Aquila, L’Aquila,  Italy.}

\author{Stefano Giagu}
\affiliation{Dipartimento di Fisica, Sapienza Università di Roma, Roma,  Italy.}

\author{Tullio Scopigno}
\affiliation{Dipartimento di Fisica, Sapienza Università di Roma, Roma,  Italy.}
\affiliation{Istituto Italiano di Tecnologia, Center for Life Nano Science @Sapienza, Roma, Italy.}
\affiliation{Istituto Italiano di Tecnologia, Graphene Labs, Genova, Italy.}

\maketitle

\section{Generation of the simulated datasets}
\subsection{Modeling of the nonlinear signals and noise fluctuations}
\begin{figure}[htbp]
\centering
\fbox{\includegraphics[width=0.95\linewidth]{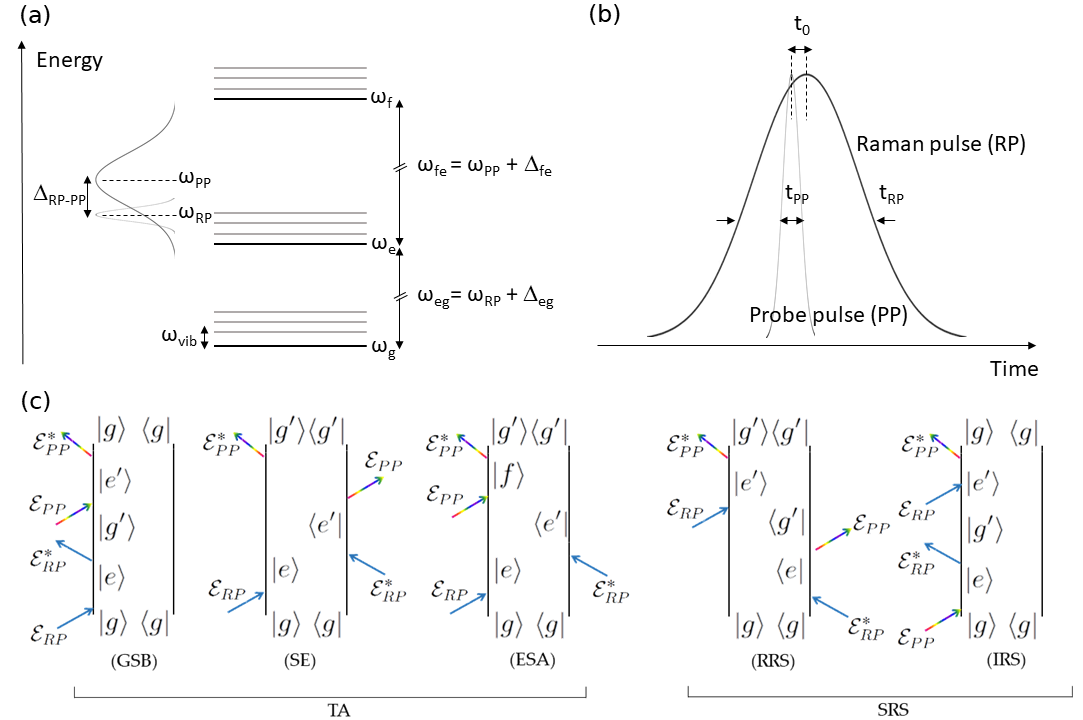}}
\caption{Sketch of the level scheme (a) and temporal envelopes of the RP and PP pulses (b). A graphical description of the main parameters used as an input in the simulations is included to clarify their physical meaning. For a full list of the parameters values see table \ref{tab:datasets}. (c) Feynman diagrams of the main contributions to the TA and SRS signals.}
\label{fig:levelscheme}
\end{figure}

The derivation of third order polarizations responsible for the SRS and TA spectroscopic signals has been reported elsewhere (see for example \cite{Mukamel_book,Fumero2015}). Briefly, nonlinear optical signals can be calculated perturbatively in the framework of the nonlinear response theory \cite{Mukamel_book}. In this formalism, the spectroscopic signal is calculated by a semiclassical procedure in which the material is treated quantum-mechanically and interacts with classical electromagnetic fields. The microscopic nonlinear polarization induced in the material is derived using the Liouville equation for the density matrix $\frac{d\rho}{d t}= - \frac{i}{\hbar}\left[H,\rho\right]$ and expanding it to the third order in the radiation-matter interaction. Diagrammatic representations are exploited to isolate the relevant terms in the perturbative expansion. Both the SRS and TA effects result from third order contributions to the nonlinear polarization. We consider the model system in Fig. \ref{fig:levelscheme}a, consisting in three electronic levels with their vibrational manifolds. The nonlinear process is described by the Hamiltonian
\begin{equation}
H=H_0+H'
\end{equation}
where $H_0$ is the free molecule and $H'$ is the dipole interaction Hamiltonian
\begin{equation}
H'(t)=-\mu\cdot \E^*(t) + c.c.
\end{equation}
where $\mu$ is the dipole operator, with matrix elements $\mu_{ij}$, and $\E(t)$ includes both the temporal envelope and the carrier of the positive frequency component of the total field 	
\begin{equation}
E(\mathbf{r},t)= \E_{RP}(t)e^{i\mathbf{k_{RP}}\cdot \mathbf{r}}+\E_{PP}(t-t_0)e^{i\mathbf{k_{PP}}\cdot \mathbf{r}}+c.c.
\end{equation}
In the pump probe scheme, SRS and TA are automatically self-matched and the signal can be derived by solving the temporal scalar problem. The $n$-th order nonlinear polarization is given by the expectation value of the dipole operator $P^{(n)}=\langle \mu \rho^{(n)}(t)\rangle = Tr(\rho^{(3)}(t))$, being $\rho^{(n)}(t)$ the density matrix evolved using the $n$-th order expansion of the Liouville equation. At the third order:
\begin{equation}
P^{(3)}(t)=\int_0^\infty d\tau_3 \int_0^\infty d\tau_2 \int_0^\infty d\tau_1 E(t-\tau_3)E(t-\tau_2-\tau_3)E(t-\tau_1-\tau_2-\tau_3)\mathcal{F}^{(3)}(\tau_1,\tau_2,\tau_3)
\end{equation}
$\mathcal{F}^{(3)}$ is the third-order matter correlation function: 
\begin{equation}\label{eq: response function}
\mathcal{F}^{(3)}(\tau_1,\tau_2,\tau_3)=\left(-\frac{i}{\hbar}\right)^3 Tr (\mu(\tau_1+\tau_2+\tau_3)[\mu(\tau_1+\tau_2)[\mu(\tau_1)][\mu(0),\rho_0]])
\end{equation}
where $\mu(t'-t'')=G^\dagger(t'-t'')\mu G(t'-t'')$ and $G$ is the retarded propagator with respect to the unperturbed Hamiltonian $H_0$:
\begin{equation}
G(t'-t'')=\theta(t'-t'') e^{-iH_0 (t'-t'')}
\end{equation}
The correlation functions can  be expanded in a sum-over-states (SoS), corresponding to the eigenfunctions of $H_0$. Each term stemming from the expansions of the commutators in eq. \ref{eq: response function} can be represented by means of Feynman diagrams, which facilitate the isolation of the relevant terms for specific signals and systems at a given order of perturbation. 
In off resonant conditions, the SRS signal is given by the sum of the two diagrams depicted in Fig. \ref{fig:levelscheme}c. In resonant conditions, additional diagrams can contribute, mainly with broad signals dominated by that of the two main diagrams \cite{Sun2008} and other pathways related to the excited state vibrational manifold \cite{Batignani2020}. In the following, we focus on ground state vibrational modes and we consider only the diagrams in Fig. \ref{fig:levelscheme}c. The results reported in the main text related to the experimental spectra of CV shows that even without including excited state manifolds explicitly in the simulated dataset, the functional dependence of the signal on the ground state diagrams is enough to disentangle the contributions of the baseline and the Raman peaks. Additional diagrams may be included to fine tune the training of the neural network for specific systems and dynamical models. The three diagrams on the left side of panel c in Fig. \ref{fig:levelscheme} account for the Ground State Bleach (GSB), Stimulated Emission (SE) and Excited State Absorption (ESA) contributing to the TA signal. The spectroscopic signal is calculated using the nonlinear polarization as a source term in the Maxwell equations. Specifically, in spectrally-resolved heterodyne detection, the signal field is given by the interferometric transmission of the PP which acts as a local oscillator:
\begin{equation}
S^{(3)}(\omega)= \mathrm{Im}\left[\E_{PP}^*(\omega)\int_{-\infty}^\infty dt\,e^{i\omega t}P^{(3)}(t)\right]
\end{equation} 
Reading off the diagrams in Fig. \ref{fig:levelscheme} \cite{Mukamel_Rahav}, we obtain:
\begin{equation}
S^{(3)}(\omega,t_0)=S_{\mathrm{RRS}}(\omega,t_0)+S_{\mathrm{IRS}}(\omega,t_0)+S_{\mathrm{GSB}}(\omega,t_0)+S_{\mathrm{SE}}(\omega,t_0)+S_{\mathrm{ESA}}(\omega,t_0)
\label{eq: signal total}
\end{equation}
where
\begin{subequations}\label{eq: signal expressions}
\begin{align}
S_{\mathrm{RRS}}(\omega,t_0)&=\sum_{e,e',g'}\frac{\mu_{ge} \mu_{eg'} \mu_{ge'} \mu_{e'g'}}{\omega-\om_{e'g'}}\E_{PP}^*(\omega,t_0)\,\int_{-\infty}^\infty d\omega''\frac{\E_{RP}(\omega'')\E_{PP}(\omega-\omega''+\omega',t_0)}{-\omega+\omega''+\om_{gg'}} \\
&\int_{-\infty}^\infty d\omega'  \frac{\E_{RP}^*(\omega')}{\omega'+\om_{ge}} \\
S_{\mathrm{IRS}}(\omega,t_0)&=\sum_{e,e',f}\frac{\mu_{ge} \mu_{eg'} \mu_{g'e'} \mu_{e'g} }{\omega-\om_{eg}}\E_{PP}^*(\omega,t_0)\,\int_{-\infty}^\infty d\omega'' \frac{\E_{RP}^*(\omega'') \E_{PP}(\omega-\omega'+\omega'',t_0)}{ \omega+\omega''-\omega'-\om_{e'g}} \\
&\int_{-\infty}^\infty d\omega'  \frac{\E_{RP}(\omega')}{\omega-\omega'-\om_{g'g}} \\
S_{\mathrm{GSB}}(\omega,t_0)&=\sum_{e,e',g'}\mu_{ge} \mu_{eg'} \mu_{g'e'} \mu_{e'g} \,\frac{|\E_{PP}(\omega,t_0)|^2}{\omega-\om_{eg}}\,\int_{-\infty}^\infty d\omega' \frac{|\E_{RP}(\omega')|^2 }{\omega'-\om_{eg}} \\
S_{\mathrm{SE}}(\omega,t_0)&=\sum_{e,e',g'}\mu_{ge} \mu_{ge'} \mu_{e'g'} \mu_{eg'} \,\frac{\E_{PP}(\omega-\omega_{ee'},t_0)\E_{PP}^*(\omega,t_0)}{\omega-\om_{eg'}}\,\int_{-\infty}^\infty d\omega' \frac{\E_{RP}(\omega')\E_{RP}^*(\omega'-\omega_{ee'}) }{\omega'-\om_{eg}} \\
S_{\mathrm{ESA}}(\omega,t_0)&=\sum_{e,e',f}\mu_{ge} \mu_{ge'} \mu_{ef} \mu_{fe'} \,\frac{\E_{PP}(\omega-\omega_{ee'},t_0)\E_{PP}^*(\omega,t_0)}{\omega-\om_{fe'}}\,\int_{-\infty}^\infty d\omega' \frac{\E_{RP}(\omega')\E_{RP}^*(\omega'-\omega_{ee'}) }{\omega'-\om_{eg}} 
\end{align}
\end{subequations}
and $t_0$ is the delay between the RP and PP and $\om_{ij}=\omega_i-\omega_j-i\gamma_{ij}$. $\gamma_{ij}$ accounts for the electronic/vibrational dephasing or for the population lifetime, depending on the state $(i,j)$.  In the simulations, the pulse envelopes have been taken as Gaussian centered at the carrier frequencies and the other parameters in eq. \ref{eq: signal expressions} are varied as specified in the next section. 

These expressions can be directly compared to the Raman Gain (RG) obtained from the experimental measurements:
\begin{equation}
\mathrm{RG}(\omega)=\frac{\Delta I_{PP}(\omega)}{I_{PP}^0(\omega)}=\frac{I_{PP}(\omega)-I_{PP}^0(\omega)}{I_{PP}^0(\omega)}
\end{equation}
where $I_{PP}$ ($I_{PP}^{0}$) is the PP spectrum detected after the sample with (without) the presence of the RP. Given the expression for the RG, we can calculate the fluctuations on the detected signal due to the shot noise using the propagation of the mean squared error. In experimental settings, the amount of noise $\mathcal{N}$ in the raw data is mitigated through averaging by a multiplicative factor which is inversely proportional to the squared root of the number of acquisitions $\mathrm{N_{acquisition}}$:
\begin{equation}
S_{\mathrm{noisy}}=\mean{S}+\frac{\mathcal{N}}{\sqrt{\mathrm{N_{acquisition}}}}
\end{equation}
$\mean{S}$ is the mean value of the distribution of obtained signals over a large number of acquisition and we take $\mathcal{N}$ as a Gaussian random variable with zero mean and standard deviation $\sigma_S$:
\begin{equation}
\sigma_S=\sigma_{\frac{\Delta I}{I_0}} =\sigma_{\frac{\Delta N}{N_0}}
\end{equation}
where we have approximated the errors on the detected intensity with those of the detected photon counts $N$ (in presence of the pump) and $N_0$ (without the pump). We assume $N$ and $N_0$ to be Poissonian, i.e. $\sigma_{N}=\sqrt{N}$ and $\sigma_{N_0}=\sqrt{N_0}$, and neglect any correlation between subsequent pulses (i.e. we assume zero covariance $\sigma_{N\,N_0}=0$) to obtain:
\begin{equation}
\sigma_S = \bigg|\frac{N}{N_0}\bigg| \,\sqrt{\frac{N}{{N_0}^2}+\frac{N^2}{{N_0}^3}}
\end{equation}
We simulated the signal on a discrete grid of frequencies $\omega_i$ with $i=1, \dots, 801$. For each spectral binning $\omega_i$, we simulated the number of signal photons in presence and absence of the RP as
\begin{equation}
N_0(\omega_i)=N_{\mathrm{counts}} \,|\E_{PP}(\omega_i)|^2 \frac{S_{\mathrm{GSB}}(\omega_i,t_0)}{\mathrm{max}(S_{\mathrm{GSB}}(\omega_i,t_0))}+1
\end{equation}
and 
\begin{equation}
N(\omega_i)=N_0 \,\mathrm{OD}\, \frac{S^{(3)}(\omega_i,t_0)}{\mathrm{max}(S (\omega_i,t_0))}+1
\end{equation}
where $N_{\mathrm{counts}}$ are the photon counts revealed by the detector, $\mathrm{OD}$ is the optical density of the material which can be used to accounts for the linear absorption of the PP (here it is set to one) and $S^{(3)}$ is given by eq. \ref{eq: signal total}.
\subsection{Parameters selection for the high and low noise datasets}

The HN and LN datasets used to train the neural networks have been generated using the model presented in the previous section. The experimental and molecular parameters were randomly extracted by uniform distributions whose limiting values are reported in table \ref{tab:datasets}. The dipole elements $\mu_{ij}$ in eq.  \ref{eq: signal expressions} have been grouped in common terms, leading to three effective transition factors that give the relative probability between TA and SRS transitions ($\mu_{TA}$), between SE/GSB and ESA transitions ($\mu_{ESA}$), and among different Raman peaks ($\mu_{vib}$). The latter are indicated in the dipole matrix elements in the SoS expressions by the primed mute indexes.  Additionally, we accounted for a different value for the number of averaged acquisitions per sample $\mathrm{N_{acquisition}}$ in each dataset, i.e. each sample in a dataset resulted from the average of $\mathrm{N_{acquisition}}$ noisy replica of the experiment. $\mathrm{N_{acquisition}}$ was set to 1 and to 100 for the HN and LN datasets, respectively. 
\begin{table}[htbp]
\centering
\resizebox{\textwidth}{!}{%
\begin{tabular}{llL}
\hline
     Parameter (units)         & Description & \mathrm{Value} \\ \hline
N\textsubscript{lim} & Number of Raman modes         &    [1,6]        \\
\textomega \textsubscript{RP} (cm\textsuperscript{-1}) & RP carrier frequency            &      20000       \\
 t\textsubscript{0} (ps) &   Time delay between the RP and PP           &  [-0.1, 0.1]            \\
 t\textsubscript{PP} (ps) &   Time duration of the PP  (FWHM)         &  0.01           \\
 t\textsubscript{RP} (ps) &   Time duration of the RP   (FWHM)        &  [3, 8]           \\
 \textDelta \textsubscript{eg} (cm\textsuperscript{-1}) &   Energy detuning between the RP and the lower electronically excited state            &  [-200, 1500]          \\
 \textDelta \textsubscript{fe} (cm\textsuperscript{-1}) &   Energy detuning between the PP and the upper electronically excited state            &  [-500, 500]          \\
\textDelta \textsubscript{PP-RP} (cm\textsuperscript{-1}) &   Energy detuning between the PP and RP      &  [-100, 100]          \\
 \textDelta \textsubscript{Stokes} (cm\textsuperscript{-1}) &  Stokes shift            &  [-200, 1500]          \\
 \textomega \textsubscript{vib} (cm\textsuperscript{-1}) & Vibrational frequencies     &       [500, 1000]      \\

\textgamma \textsubscript{vib} (ps) & Vibrational dephasing time    &       [0.1,3]      \\
\textGamma (ps) & Excited-state population lifetime    &        [50,500]     \\
\textgamma \textsubscript{e} (cm\textsuperscript{-1}) & Electronic dephasing rate   &        [800,2000]     \\
\textmu \textsubscript{vib}  & Relative intensity of different Raman transitions    &        [0.25, 1]     \\
 \textmu \textsubscript{TA}  & Relative intensity between Raman and TA interactions   &        [1,7]     \\
 \textmu \textsubscript{ESA}  & Relative intensity between ESA and SE/GSB interactions   &        [0.5,5]     \\
 
N\textsubscript{counts}  & Photon counts detected by the detector   &        [5,60] \cdot 10^3     \\

N\textsubscript{acquisition}  & Number of averaged acquisitions   &        \text{High noise dataset: } 1     \\
&   &        \text{Low noise dataset: } 100     \\

 \\ \hline  
\end{tabular}%
}
\caption{Experimental and molecular parameters used to generate the simulated datasets. Values in brackets refer to the lower and upper limits of the uniform distribution from which the actual parameters are extracted for each sample of the dataset. $\omega_{RP}$ and $\mathrm{t_{PP}}$, for which single values are shown in the table, were kept fixed for all the samples. We refer to Fig. \ref{fig:levelscheme} for a graphical description of the physical meaning of the parameters.}
\label{tab:datasets}
\end{table}
\section{Analysis of the performance of HN and LN networks on the simulated datasets}
\subsection{Additional analysis of the performances of the HN network}
\begin{figure}[htbp]
\centering
\fbox{\includegraphics[width=0.95\linewidth]{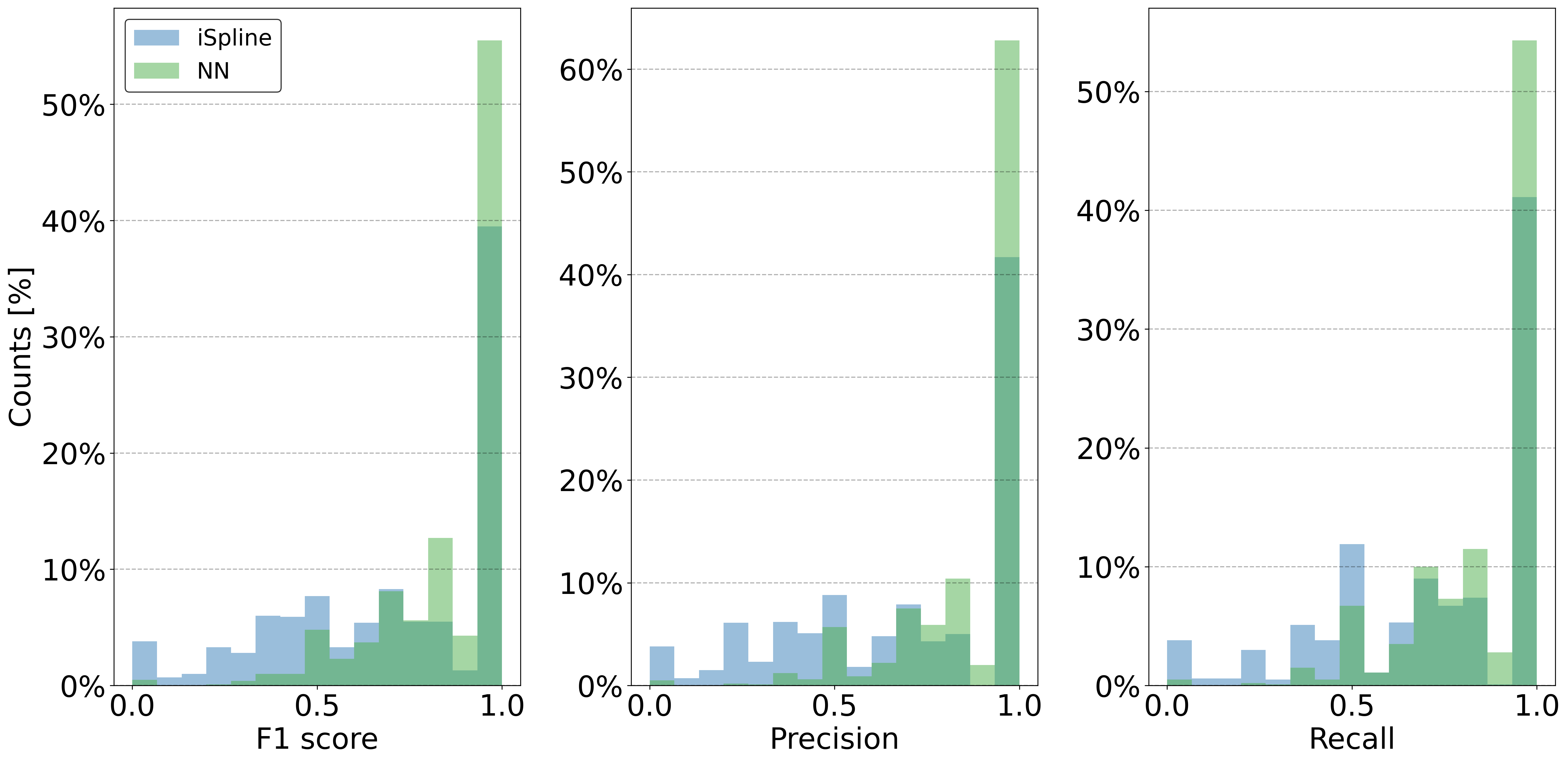}}
\caption{Histograms of the F1 score, precision and recall obtained on the test set by the HN network and the iSpline algorithm for baseline removal followed by a SG noise filter.}

\label{fig:hist_classification_HN}
\end{figure}

\begin{figure}[htbp]
\centering
\fbox{\includegraphics[width=0.95\linewidth]{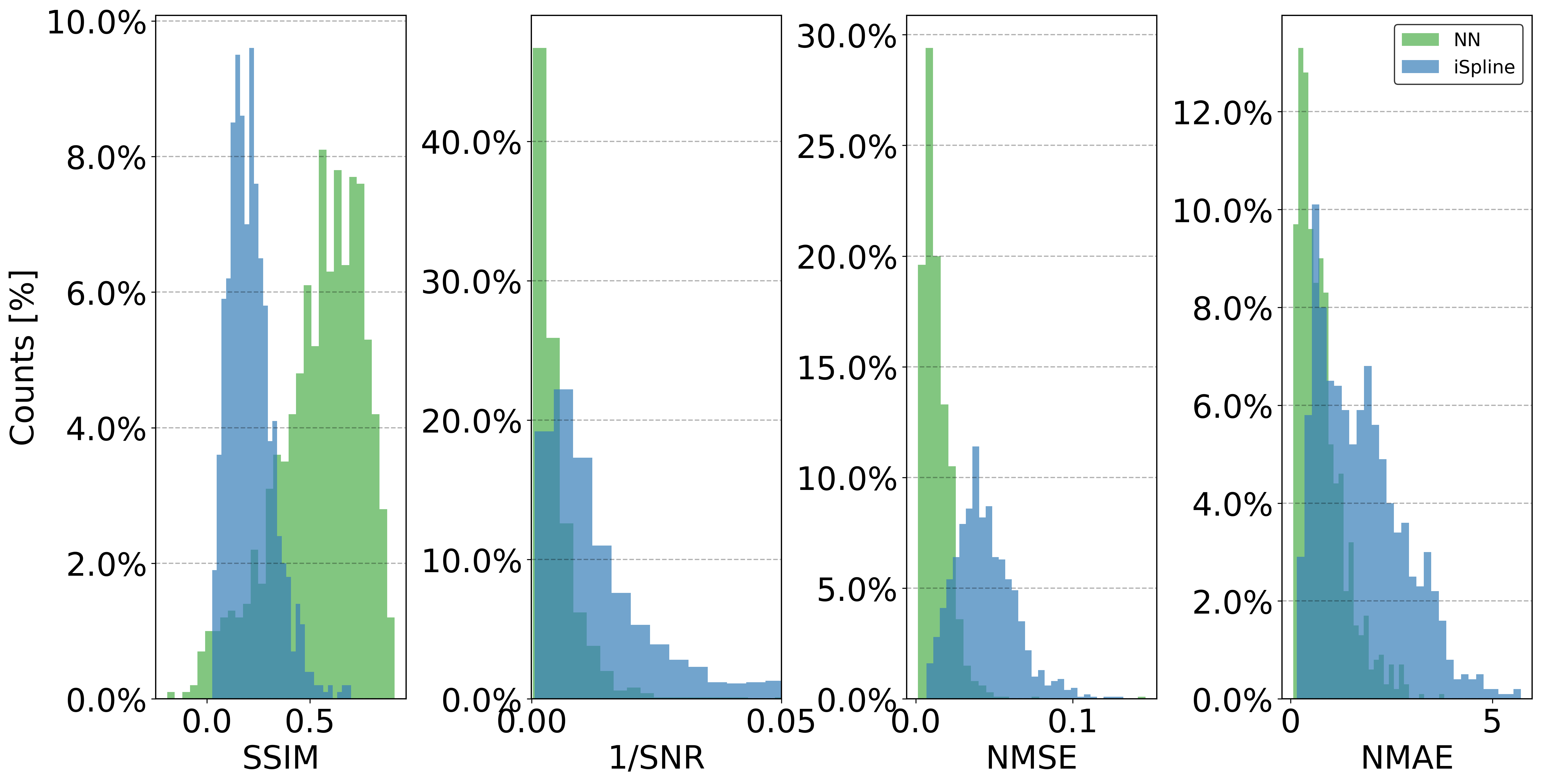}}
\caption{Histograms of the performances obtained for selected metrics on the test set by the HN network and the iSpline algorithm for baseline removal followed by a SG noise filter.}
\label{fig:hist_metrics_HN}
\end{figure}

In Fig. \ref{fig:hist_classification_HN}-\ref{fig:hist_metrics_HN}, we report the histograms of the classification and regression metrics for the HN network and the iSpline method, trained and tested on independent subsets of the HN dataset.

\begin{figure}[htbp]
\centering
\fbox{\includegraphics[width=0.95\linewidth]{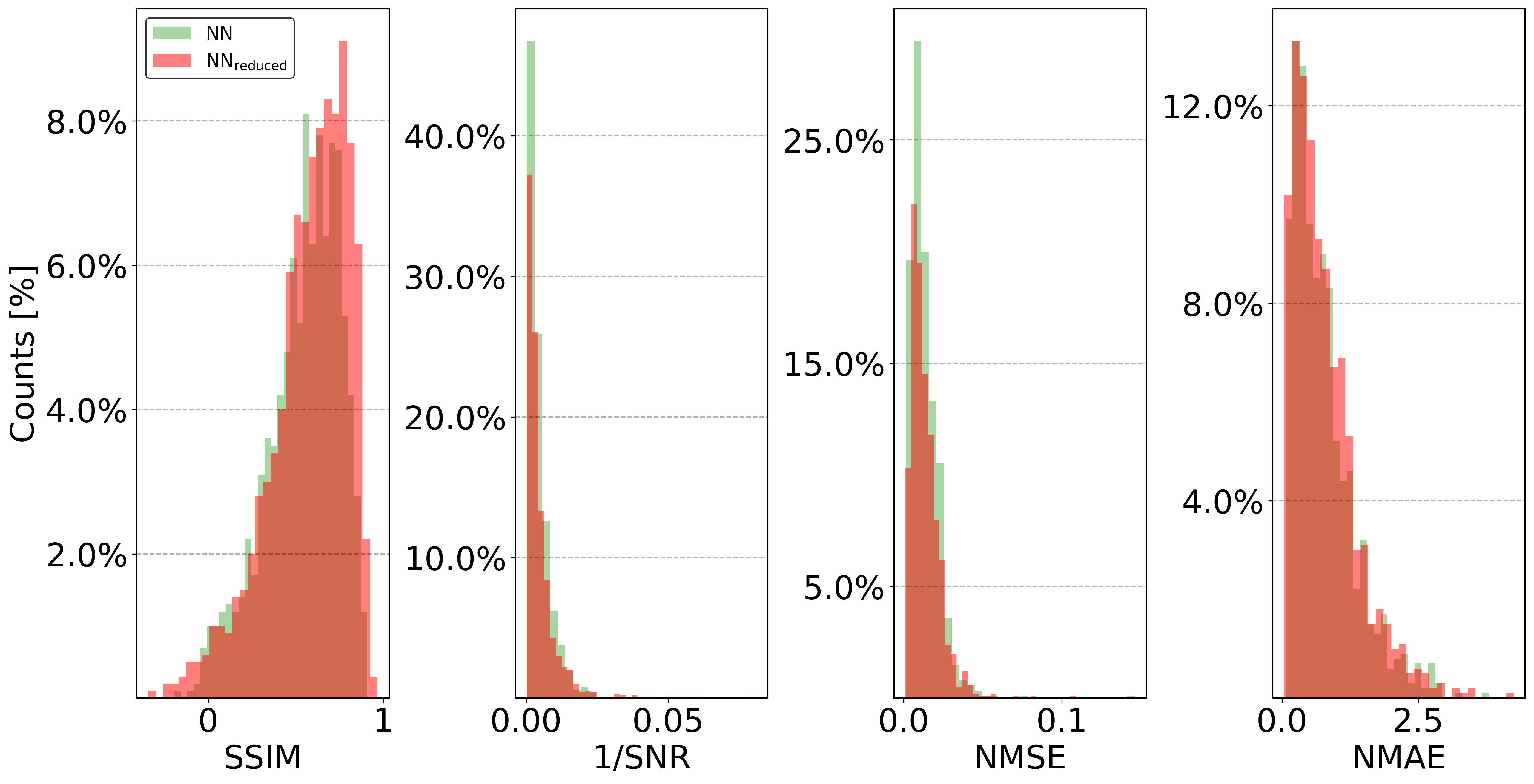}}
\caption{Histograms of the performances obtained for selected metrics on the test set by the HN network and its reduced version.}
\label{fig:metrics_NNvsNN2}
\end{figure}

\begin{figure}[htbp]
\centering
\fbox{\includegraphics[width=0.95\linewidth]{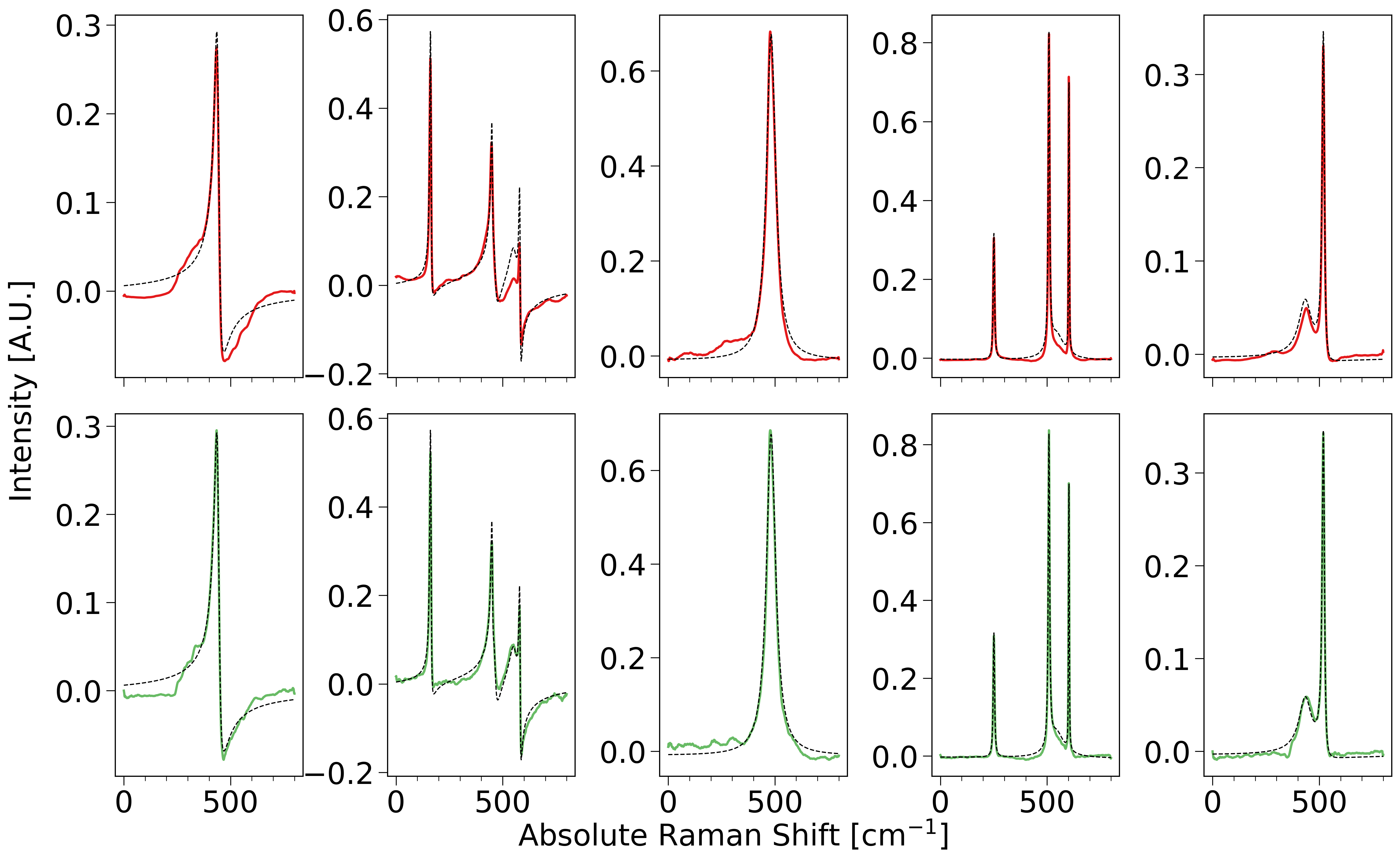}}
\caption{Comparison between of evaluation of the selected HN test samples by the NN (bottom panel, green) and NN\textsubscript{reduced} (top panel, red). The samples and the output of the NN network are the same of those reported in Fig. 3 of the main text. GT are also reported for comparison (black dotted lines).}
\label{fig:qualitative_NNvsNN2}
\end{figure}

We now evaluate the dependence of the HN network on the number of parameters in the convolutional branches. $N_{param}$ and $N_{kernel}$ define the majority of the parameters of the network and thus largely determines its complexity and computational cost for training. Their values are larger for the HN network with respect to the LN.  We found that similar results for the metrics can be obtained on the HN dataset using a network with $N_{param}=10k$ and $N_{kernel}=21$, the same number of parameters used for the LN, by using ${\ell}=1$ norm for the gradient component of the loss function and re-training the network. This is shown in Fig. \ref{fig:metrics_NNvsNN2}, where NN\textsubscript{reduced} is the new, lighter network with 10k parameters in the convolutional branches. This may be indicative of the fact that a further increase of the number of parameters, besides requiring more computational cost, does not necessarily guarantee an increase of the performances. However, a closer look at specific samples in the dataset demonstrates that NN\textsubscript{reduced} is less effective in recovering the details of the lineshape in situations with large noise (see for example the relative intensities of the peaks in panel 2 and the shoulder of the central peak in panel 4 in Fig. \ref{fig:qualitative_NNvsNN2}). The analysis is also indicative of the necessity to find additional metrics to make the evaluation of the performance more robust against subtle differences between GT and reconstructed lineshapes. 

\subsection{Dependence of the classification metrics on the tolerance of the edge finder}
\begin{figure}[htbp]
\centering
\fbox{\includegraphics[width=0.95\linewidth]{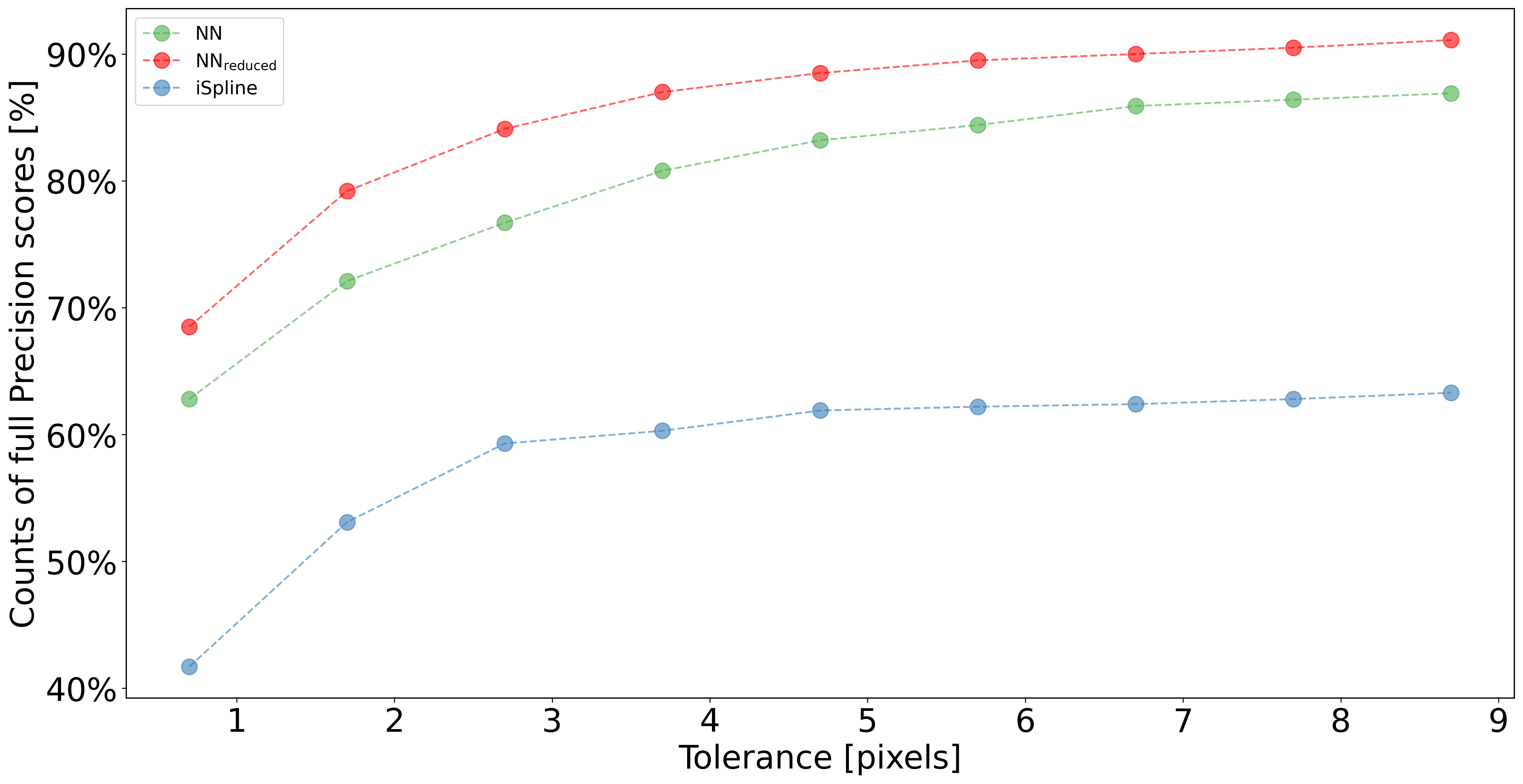}}
\caption{Precision score on the HN test set as a function of the tolerance of the edge finder for the NN methods and iSpline.}
\label{fig:metrics_toler}
\end{figure}

The value of classification metrics based on the edge finder for a specific algorithm depends on the amount of tolerance on the peak position discrepancy between the GT and reconstructed data. Setting a very small tolerance (as in the case of the analysis reported in Fig. 3a) reports on the performance of the algorithm in recovering both the peaks positions and the number of peaks. A larger value of tolerance is less sensitive to the peak position and consequently allows to discriminate the cases in which the number of peaks is not correctly retrieved. These dependence affects similarly the different algorithms and does not affect the analysis of the performances reported in the main text. This is shown in Fig. \ref{fig:metrics_toler}, in which we report the percentage of perfect score on precision (precision=1) obtained by different algorithms as a function of the tolerance in pixels. For each value of the tolerance, the best performances are obtained by the NN and NN\textsubscript{reduced}. The best standard algorithm achieves a full score for about $\mathrm{20\%}$ fewer samples than the threshold set by neural networks. 

\subsection{Results on the LN dataset}
\begin{figure}[htbp]
\centering
\fbox{\includegraphics[width=0.95\linewidth]{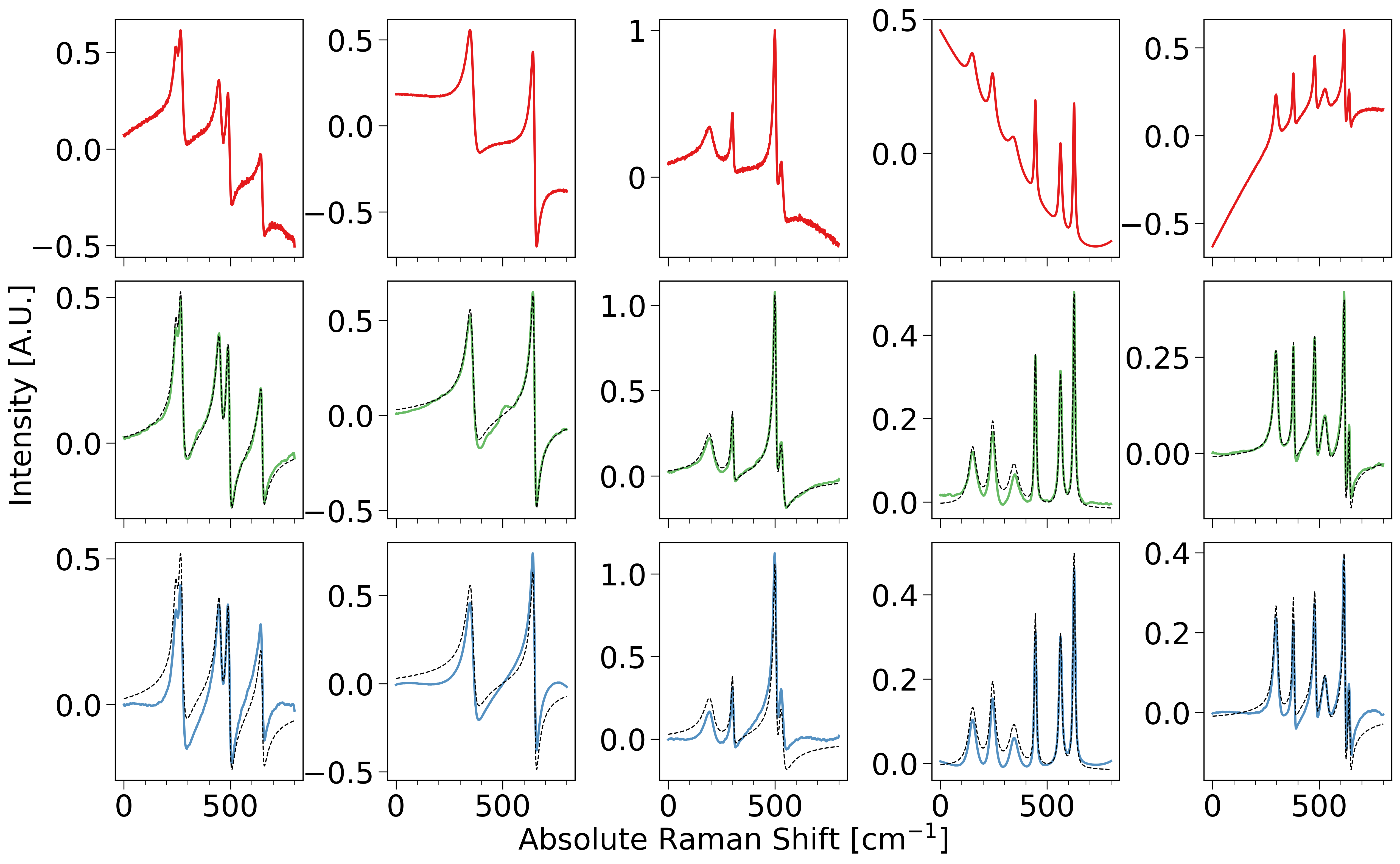}}
\caption{Examples of evaluation of the simulated test samples from the LN dataset. For each sample, noisy raw data are shown in the top panel (red lines). Central and bottom panels show the SRS spectra obtained by using the best NN model (green) or the iterative spline procedure for baseline subtraction followed by a Savitzky-Golay filtering (blue). GT are also reported for comparison (black dotted lines).}
\label{fig:simulated_spectra_LN}
\end{figure}

\begin{figure}[htbp]
\centering
\fbox{\includegraphics[width=0.95\linewidth]{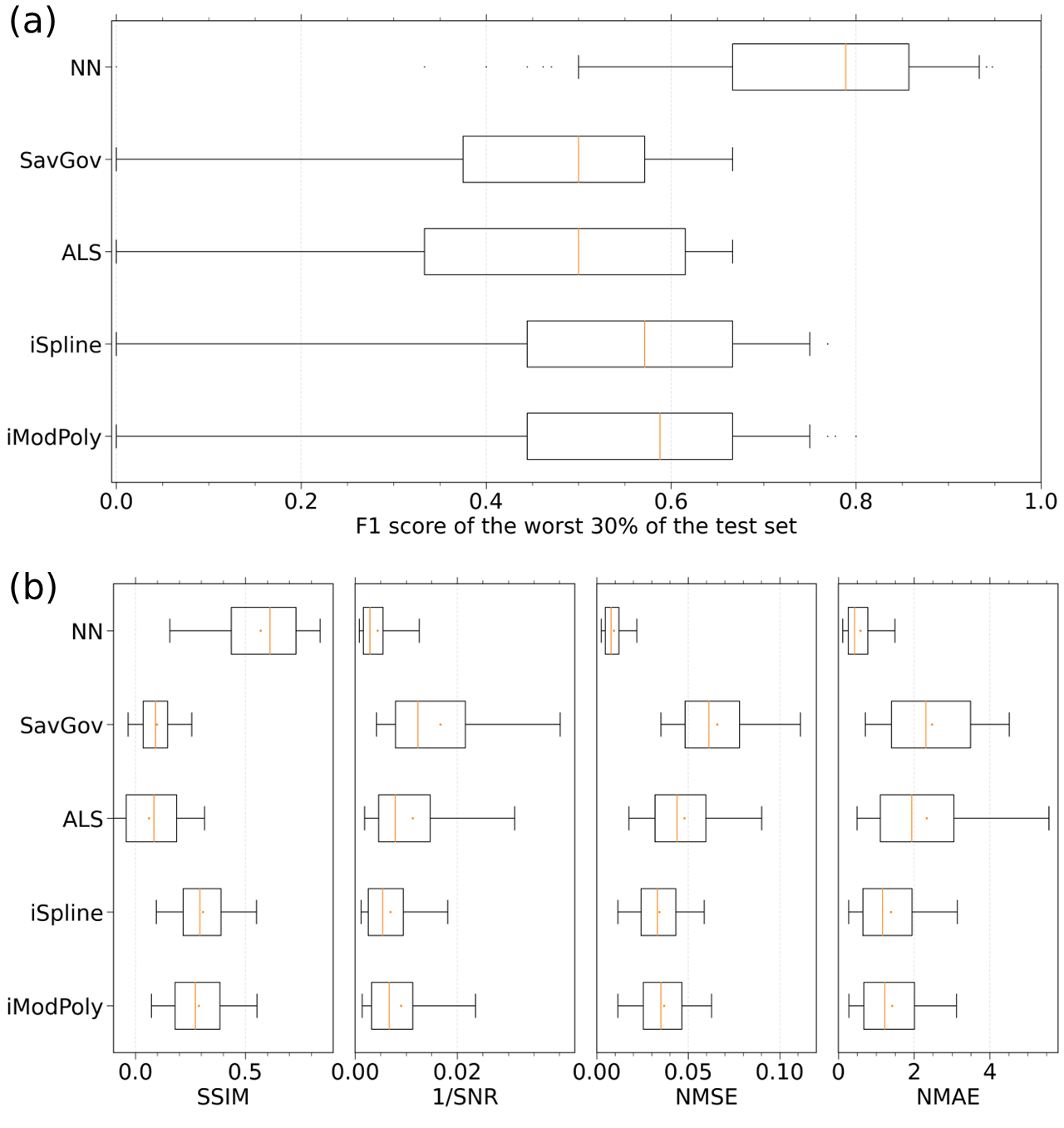}}
\caption{Comparison of the statistical analysis over selected metrics between the NN and multiple standard routines applied to the LN test set. Panel a shows the results of the classification metric obtained by means of the edge finder algorithm. The whisker plots report on the F1 score relative to the portion of 30\% of the test set for which each method has the worst score. Panel b shows the whisker plots relative to the results obtained by the different methods measure by the SSIM, SNR, NMSE and NMAE metrics. For all the whisker plots, the box covers from first to the third quartile, while the whiskers extend from the box to the the 5th and to the 95th percentile. The orange line and dot indicate the median and the mean, respectively. Black dots indicate values that are past the end of the whiskers.}
\label{fig:metrics_LN}
\end{figure}

\begin{figure}[htbp]
\centering
\fbox{\includegraphics[width=0.95\linewidth]{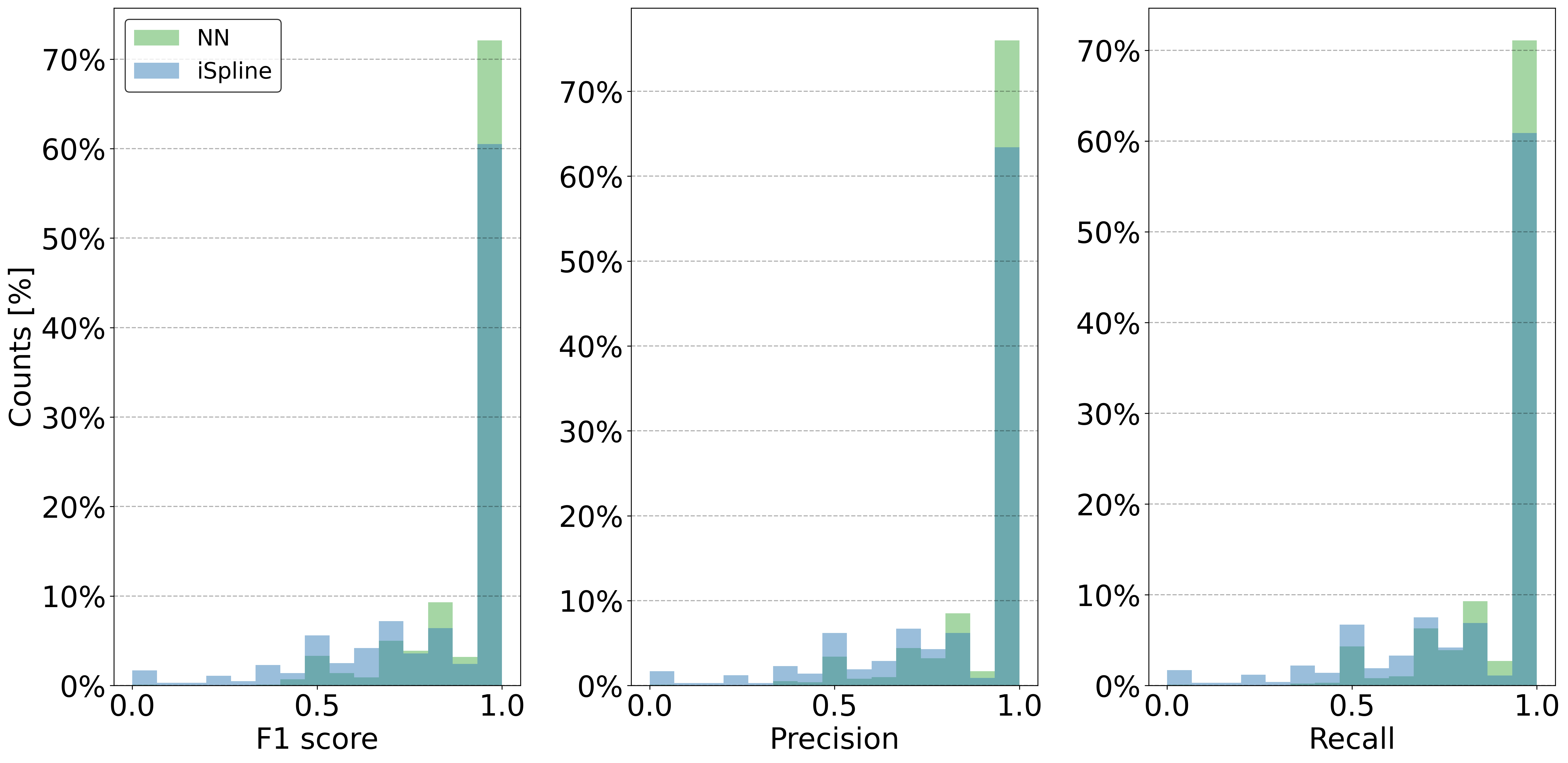}}
\caption{Histograms of the F1 score, precision and recall obtained on the test set by the LN network and the iSpline algorithm for baseline removal followed by a SG noise filter.}
\label{fig:hist_classification_LN}
\end{figure}

\begin{figure}[htbp]
\centering
\fbox{\includegraphics[width=0.95\linewidth]{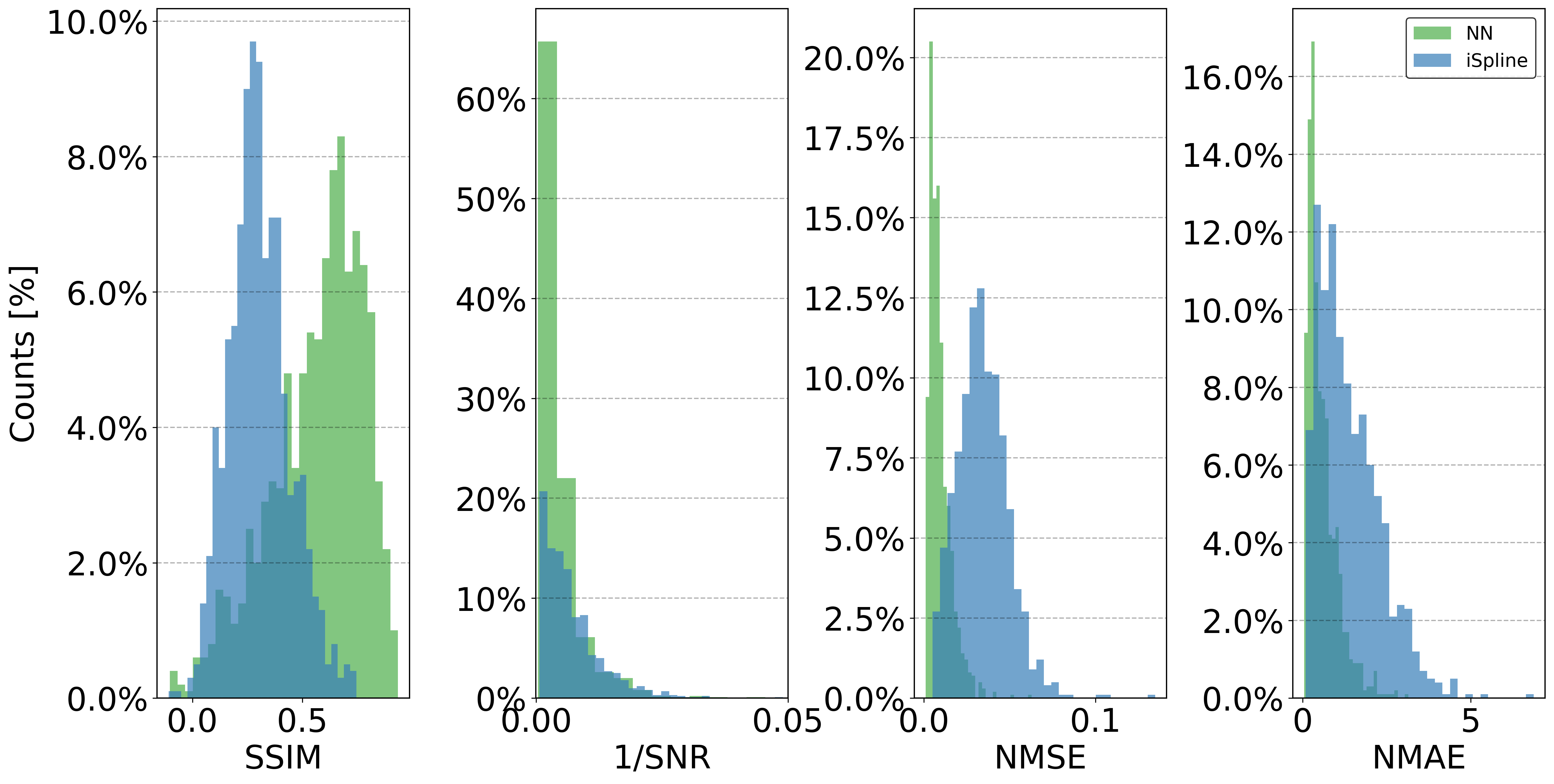}}
\caption{Histograms of the performances obtained for selected metrics on the test set by the LN network and the iSpline algorithm for baseline removal followed by a SG noise filter.}
\label{fig:hist_metrics_LN}
\end{figure}

In this section, we report the analysis of the performance of the LN network, trained and tested on independent subsets of the LN dataset. In Fig. \ref{fig:simulated_spectra_LN}, we report the reconstruction of clean lineshapes for typical samples in the LN test set and compared the results of the LN network with those of the iSpline algorithm followed by SavGov filtering.
A comparison between the performance of the NN and the standard algorithms is reported in Fig. \ref{fig:metrics_LN}. Similarly to what observed for the HN dataset, the NN outperforms the standard methods for all the tested metrics. In Fig. \ref{fig:hist_classification_LN}-\ref{fig:hist_metrics_LN}, we report the histograms of the classification and regression metrics for the LN network and the iSpline method. 

\subsection{Combining different networks for different noise level} \label{sec: combining NN}
The HN and LN network can be combined to tackle input data with an unknown level of noise by means of a discriminator. We first obtained a raw estimate of the SNR using the following procedure, which is applicable also to experimental data. We  extracted $n$ random spectra from the simulated dataset and select a narrow spectral region in which Raman peaks are not present. Then we subtracted a raw local baseline, which is well approximated by applying a moving average filter and calculate the standard deviation. This procedure gives an estimation of the SNR, whose error can be inferred by changing the extrema of the spectral region and the number of spectra and repeating the calculation. For the two simulated dataset, we obtained $\sigma_{high}=1e^{-2}$ and $\sigma_{low}=1e^{-3}$. The choice of the NN architecture can be adapted to the noise of the experimental data using a discriminator which measures $\sigma$ for the input and then adopt one of the two networks depending on the result: for $\sigma$ greater than a threshold fixed at $0.5e^{-3}$, the input is processed by the network LN, otherwise by the network HN. Alternatively, given the adaptability of the parallel kernel convolutional architecture on very different amount of noise, depending on the details of the experimental setups and data acquisitions, the NN can be retrained or selectively fine tuned on specific layers, including more noise levels and additional sources of noise.

\section{Frequencies of the Raman modes in the experimental spectra}

\subsection{Deoxy myoglobin}
\begin{table}[H]
\centering
\begin{tabular}{c}
\hline
     Raman mode frequency (\icm)         \\ \hline 
   1115 \\ 1216 \\1353\\1428\\1446\\1467\\1520\\1550\\1561\\1587\\1616     \\
 \hline  
\end{tabular}%
\caption{Spectral positions of the Raman modes of deoxy Myoglobin in the Soret band from literature \cite{Ferrante2018}, reported as black vertical lines in Fig. 5a of the main text. }
\label{tab:mb}
\end{table}
\subsection{Cresyl Violet}
\begin{table}[H]
\centering
\begin{tabular}{|c|c|c|c|}
\hline
     \multicolumn{2}{|c|} {Ground state modes}        & \multicolumn{2}{c|} {Excited state modes}   \\ \hline
     Frequency (\icm) & Ref. &  Frequency (\icm) & Ref.
     \\ \hline
     
   347 &  \cite{Lu2020}      &    342  & \cite{Fitzpatrick2020,Lu2020}           \\
   472 &  \cite{Lu2020}          &      469 & \cite{Batignani2020}            \\

   494 & \cite{Batignani2020}      &       489 & \cite{Fitzpatrick2020,Batignani2020}         \\
   524 & \cite{Batignani2020}       &  526 &  \cite{Batignani2020}           \\
   580 &\cite{Batignani2020} & 570 & \cite{Fitzpatrick2020,Batignani2020}   \\
   592 &  \cite{Fitzpatrick2020,Batignani2020, Batignani2021}   & 589 & \cite{Fitzpatrick2020,Batignani2020} \\
   674 & \cite{Batignani2020}  & 669 & \cite{Batignani2020} \\
  739 & \cite{Fitzpatrick2020}       &       726  & \cite{Batignani2020}      \\
     822 & \cite{Fitzpatrick2020}   & 821 & \cite{Fitzpatrick2020}\\ \hline  
\end{tabular}%
\caption{Spectral positions of the ground and excited states Raman modes in CV from literature, reported as black and orange vertical lines in Fig. 5b of the main text. }
\label{tab:cv}
\end{table}


%